\documentclass[aps,prd,12pt,superscriptaddress,eqsecnum,amsmath,nofootinbib,preprintnumbers,longbibliography]{revtex4-2}%

\usepackage{color,graphicx,float}
\usepackage{amsfonts,amssymb,theorem,mathrsfs,times}
\usepackage{bm}
\usepackage{mathtools}
\usepackage{dsfont}
\usepackage{setspace}
\usepackage[normalem]{ulem}
\usepackage[colorlinks,linkcolor=blue,anchorcolor=blue,citecolor=blue]{hyperref}
\usepackage{orcidlink}
\usepackage{comment}
\usepackage{subcaption}
\usepackage{caption}
\usepackage{cancel}

{\theorembodyfont{\upshape}
}
{\theorembodyfont{\upshape}
}
{\theorembodyfont{\upshape}
}
{\theorembodyfont{\upshape}
}
{\theorembodyfont{\upshape}
}
{\theorembodyfont{\upshape}
}

\newcommand{\dalm}{\kern1pt\vbox{\hrule height 0.9pt\hbox{\vrule width
0.9pt\hskip 2.5pt\vbox{\vskip 5.5pt}\hskip 3pt\vrule width
0.3pt}\hrule height 0.3pt}\kern1pt}

\begin{document}

\title{Black holes with dark matter halos and a positive cosmological constant: Energy conditions and signatures}

\author{Long-Yue Li  \orcidlink{0000-0002-7785-1116}}
\email{lilongyue@nbu.edu.cn}
\affiliation{Zhejiang Key Laboratory of Extreme Universe, Institute of Fundamental Physics and Quantum Technology, Department of Physics, School of Physical Science and Technology, Ningbo University, Ningbo, Zhejiang 315211, China}

\author{Li-Ming Cao  \orcidlink{0000-0001-6633-6613}}
\email{caolm@ustc.edu.cn}
\affiliation{
Interdisciplinary Center for Theoretical Study and Department of Modern Physics,\\
University of Science and Technology of China, Hefei, Anhui 230026, China}

\author{Yungui Gong \orcidlink{0000-0001-5065-2259}}
\email{gongyungui@nbu.edu.cn}
\affiliation{Zhejiang Key Laboratory of Extreme Universe, Institute of Fundamental Physics and Quantum Technology, Department of Physics, School of Physical Science and Technology, Ningbo University, Ningbo, Zhejiang 315211, China}

\author{Xia-Yuan Liu  \orcidlink{0009-0008-3852-0147}}
\email[corresponding author: ]{liuxiayuan@mail.ustc.edu.cn}
\affiliation{
Interdisciplinary Center for Theoretical Study and Department of Modern Physics,\\
University of Science and Technology of China, Hefei, Anhui 230026, China}

\author{Liang-Bi Wu  \orcidlink{0000-0003-0107-4944}}
\email{wulb@ucas.ac.cn}
\affiliation{School of Fundamental Physics and Mathematical Sciences, Hangzhou Institute for Advanced Study, UCAS, Hangzhou, Zhejiang 310024, China}

\author{Wenting Zhou  \orcidlink{0000-0003-4046-753X}}
\email{zhouwenting@nbu.edu.cn}
\affiliation{Zhejiang Key Laboratory of Extreme Universe, Institute of Fundamental Physics and Quantum Technology, Department of Physics, School of Physical Science and Technology, Ningbo University, Ningbo, Zhejiang 315211, China}

\begin{abstract}
We study static, spherically symmetric black holes surrounded by dark matter halos in the presence of a positive cosmological constant. Assuming vanishing radial pressure, we derive the constraints imposed by the null, weak, strong, and dominant energy conditions on the halo matter. The halo is confined to a finite radial interval separated from both horizons, and the total mass is bounded above by the Nariai value. 
We construct a Hernquist-type profile and study black hole images, showing that a positive cosmological constant can substantially change the images.
We also compute scalar quasinormal modes of the Hernquist-type spacetime. When the dark matter mass is sufficiently small, the relative difference of a selected mode is approximately linear in the compactness of the matter halo, i.e.,  the ratio of mass parameter $M_{\text{DM}}$ and scale parameter $a$, with a coefficient that depends on the cosmological constant.
\end{abstract}

\maketitle
\raggedbottom

\section{Introduction}\label{S1}
Dark matter accounts for approximately $26.8\%$ of the cosmic energy density inferred from the Planck analysis~\cite{Planck:2018vyg}. Its gravitational effects are evident in galaxy rotation curves~\cite{Rubin:1970zza,Rubin:1978kmz} and gravitational lensing, including the separation between the lensing mass distribution and the X-ray-emitting gas in the Bullet Cluster~\cite{Clowe:2006eq,Bradac:2006er}. These observations motivate the study of dark matter halos, although the nature of their constituent particles remains unknown~\cite{Bertone:2004pz}.

Several density profiles are used to model galactic matter distributions. The Hernquist profile was introduced for spherical galaxies and bulges~\cite{Hernquist:1990be}, whereas the Navarro--Frenk--White (NFW) profile arose from cold dark matter simulations~\cite{Navarro:1994hi}. The Jaffe profile describes the light distribution in spherical galaxies~\cite{Jaffe:1983iv}. Other examples include the Burkert profile for dwarf galaxies~\cite{Burkert:1995yz,Salucci:2000ps} and the profiles considered by Taylor and Silk in studies of halo clumpiness and annihilation signals~\cite{Taylor:2002zd}.

A relativistic description is needed when a halo surrounds a black hole. Cardoso \textit{et al.} constructed a black hole spacetime with a Hernquist-type halo whose density vanishes at the event horizon and peaks outside it~\cite{Cardoso:2021wlq}. Charged and rotating extensions have since been developed~\cite{Stelea:2023yqo,Fernandes:2025osu}. Studies of these environments have examined thermodynamics~\cite{Myung:2025cxu}, quasinormal modes (QNMs)~\cite{Feng:2025iao,Pezzella:2024tkf}, gravitational waves from extreme mass-ratio inspirals~\cite{Dai:2023cft,Zhao:2024bpp,Zhao:2026yis}, and black hole images~\cite{Xavier:2023exm,Macedo:2024qky}. Analytical constructions have also been extended to other halo profiles~\cite{Shen:2023erj}.

Specifying a density profile, however, does not guarantee that the resulting energy-momentum tensor satisfies the energy conditions. The model of Ref.~\cite{Cardoso:2021wlq} violates the dominant energy condition near the event horizon.
In Newtonian mechanics and general relativity, particles near a gravitational center have different innermost distributions~\cite{Gondolo:1999ef,Sadeghian:2013laa}. Nevertheless, these results all assume test particles in a fixed background spacetime and neglect the backreaction of matter on the spacetime geometry.
For a static, spherically symmetric halo with vanishing radial pressure, imposing all four energy conditions gives an inner bound of $2.5M_{\text{B}}$ in the asymptotically flat case, where $M_{\text{B}}$ is the black hole mass~\cite{Shen:2024qbb}. This provides an energy-condition-compatible extension of the original construction.

A positive cosmological constant introduces a cosmological horizon and changes the large-scale geometry. Its role on different astrophysical scales has been investigated in Refs.~\cite{Benisty:2023clf,HassanPuttasiddappa:2026cuh}. Charged black holes with halos in a de Sitter background were considered in Ref.~\cite{Konoplya:2025nqv}, but the energy condition constraints addressed here were not analyzed there. We extend the framework of Refs.~\cite{Cardoso:2021wlq,Shen:2024qbb} to $\Lambda>0$. Under the same vanishing-radial-pressure assumption, the energy conditions confine the halo to a finite interval separated from both horizons. They also bound the total mass by the Nariai value~\cite{Nariai:1999iok}.

We construct a Hernquist-type profile within this allowed region and examine the images produced by several halo models.
Motivated by studies of black hole images~\cite{Gralla:2019xty,Peng:2020wun,Gan:2021pwu,Cao:2025qzy,Qin:2023nog}, we compare the photon effective potentials and images with and without $\Lambda$. The images differ significantly depending on whether
$\Lambda$ is included, but this alone does not allow the cosmological constant to be measured.
We also calculate scalar QNMs, tracking selected $\ell=2$ modes as $\Lambda$, the halo scale radius $a$, and the mass parameter $M_{\text{DM}}$ vary. When the dark matter mass is sufficiently small, the relative frequency shift is approximately proportional to $M_{\text{DM}}/a$, with a coefficient that depends on $\Lambda$.

This paper is organized as follows. Section~\ref{S2} derives the energy-condition constraints, and Sec.~\ref{S3} constructs the Hernquist-type spacetime. Sections~\ref{S4} and \ref{S5} present the images and scalar QNMs, respectively. Section~\ref{S6} summarizes the results and discusses extensions. 
In addition, we discuss black hole images for several other dark matter models in Appendix~\ref{S7}.

\section{Energy conditions for black holes with dark matter halos}\label{S2}
Following Refs.~\cite{Cardoso:2021wlq,Shen:2024qbb}, we consider a static, spherically symmetric halo with vanishing radial pressure. This anisotropic-fluid ansatz is motivated by the Einstein-cluster description of particles on circular orbits~\cite{Boehmer:2007az}. The halo stress tensor is

\begin{eqnarray}
    T^\mu{}_\nu=\text{diag}\Big(-\rho(r), 0, P(r), P(r) \Big)  \, .
\end{eqnarray}
where $\rho(r)$ is the energy density and $P(r)$ is the tangential pressure. The cosmological constant is treated separately from the halo stress tensor. The metric takes the form
\begin{eqnarray}
    \mathrm{d}s^2=-f(r)\mathrm{d}t^2+\dfrac{\mathrm{d}r^2}{1-\dfrac{2m(r)}{r}-\dfrac{\Lambda r^2}{3}}+r^2\mathrm{d}\Omega^2  \, .
\end{eqnarray}
The event and cosmological horizons, denoted by $r_{\text{h}}$ and $r_{\text{c}}$, are the relevant positive roots of
\begin{eqnarray}
    1-\frac{2m(r)}{r}-\frac{\Lambda r^2}{3}=0  \, .
\end{eqnarray}
We assume that dark matter is confined to the region $r>r_1$. Therefore, the region $r<r_1$ is vacuum, and $m(r)=M_{\text{B}}$, where $M_{\text{B}}$ is the mass of the black hole.
Hence, $r_{\text{h}}$ is the smaller positive root of $1-2M_{\text{B}}/r-\Lambda r^2/3=0$.
The Einstein equations give
\begin{eqnarray}
\label{fE} \frac{f'}{f}&=&\frac{2}{r^2}\frac{m-\frac{\Lambda r^3}{3}}{1-\frac{2m}{r}-\frac{\Lambda r^2}{3}}       \, ,\\
\label{rhoE} \rho&=&\frac{m'}{4\pi r^2}       \, ,\\
\label{pE} \frac{P}{\rho}&=&\frac{ \frac{m}{r}-\frac{\Lambda r^2}{3} }{2\big( 1-\frac{2m}{r}-\frac{\Lambda r^2}{3} \big)}   \, ,
\end{eqnarray}
where a prime denotes differentiation with respect to $r$. We impose the null, weak, strong, and dominant energy conditions on the halo matter. With vanishing radial pressure, their joint requirements are
\begin{eqnarray}\label{rhoP}
    \rho&\geqslant&0   \, ,\nonumber\\
    \rho+P&\geqslant&0   \, ,\nonumber\\
    \rho+2P&\geqslant&0   \, ,\nonumber\\
    \rho-P&\geqslant&0     \, .
\end{eqnarray}
In the region between the event horizon and the cosmological horizon, the condition
\begin{eqnarray}\label{glarger0}
    1-\frac{2m}{r}-\frac{\Lambda}{3}r^2\geqslant0
\end{eqnarray}
must also be satisfied. Wherever $\rho>0$, Eqs.~(\ref{pE}), (\ref{rhoP}), and (\ref{glarger0}) imply
\begin{eqnarray}
    2r-3m-\Lambda r^3&\geqslant&0   \, ,\nonumber\\
    r-m-\frac{2}{3}\Lambda r^3&\geqslant&0    \, ,\nonumber\\
    2r-5m-\frac{1}{3}\Lambda r^3&\geqslant&0      \, ,\nonumber\\
    r-2m-\frac{\Lambda}{3}r^3&\geqslant&0   \, .
\end{eqnarray}
Since $m(r)$ contains the black hole mass and the halo mass, $m(r)\geqslant M_{\text{B}}$. Thus all four upper bounds below must hold simultaneously:
\begin{eqnarray}\label{mLambda}
    M_{\text{B}}\leqslant m(r)\leqslant  \left\{
\begin{array}{ll}
    \frac{2}{3}r-\frac{1}{3}\Lambda r^3 &=:m_1(r) \,,  \\
    r-\frac{2}{3}\Lambda r^3 &=:m_2(r)  \,, \\
    \frac{2}{5}r-\frac{1}{15}\Lambda r^3 &=:m_3(r) \,, \\
    \frac{1}{2}r-\frac{1}{6}\Lambda r^3  &=:m_4(r)   \,.
\end{array}
\right.  
\end{eqnarray}

\begin{figure}[tbp]
    \centering
    \includegraphics[width=10cm]{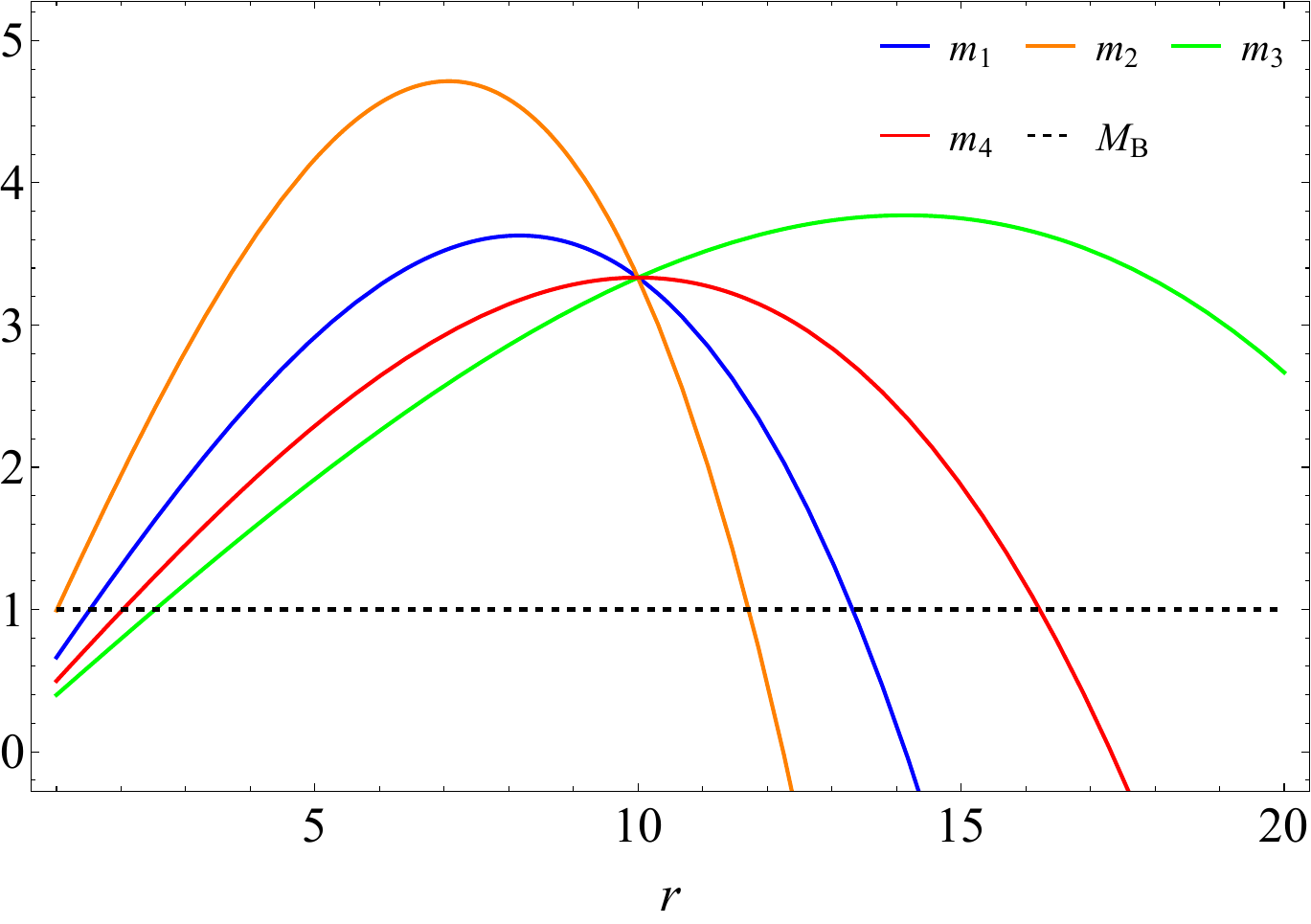}
    \caption{The four mass bounds in Eq.~(\ref{mLambda}) for $M_{\text{B}}=1$ and $\Lambda=0.01$.}
    \label{m1234}
\end{figure}

The four bounding curves meet at
\begin{eqnarray}\label{r0}
    r_0=\frac{1}{\sqrt{\Lambda}}  \, .
\end{eqnarray}
as illustrated in FIG.~\ref{m1234}. The tightest bound is $m_3(r)$ below $r_0$ and $m_2(r)$ above $r_0$, so the constraints reduce to
\begin{eqnarray}\label{mrange}
M_{\text{B}}\leqslant m(r)\leqslant  \left\{
\begin{array}{ll}
    \frac{2}{5}r-\frac{1}{15}\Lambda r^3  \, ,\qquad& r_{\text{1}}\leqslant r\leqslant\frac{1}{\sqrt{\Lambda}} \,,\\
     r-\frac{2}{3}\Lambda r^3   \, ,\qquad&  \frac{1}{\sqrt{\Lambda}}<r \leqslant r_{\text{2}} \,.
\end{array}
\right.  
\end{eqnarray}
Here
\begin{eqnarray}
    r_{1}=2\sqrt{\frac{2}{\Lambda}}\cos\Bigg[\frac{1}{3}\arccos\Big( -\frac{15M_{\text{B}}\sqrt{\Lambda}}{4\sqrt{2}}  \Big)- \frac{2\pi}{3} \Bigg]
\end{eqnarray}
is the smaller positive root of
\begin{eqnarray}
\frac{2}{5}r-\frac{1}{15}\Lambda r^3-M_{\text{B}}=0  \, ,
\end{eqnarray}
and
\begin{eqnarray}
    r_{2}=\sqrt{\frac{2}{\Lambda}}\cos\Bigg[\frac{1}{3}\arccos\Big( -\frac{3M_{\text{B}} \sqrt{2 \Lambda}}{2} \Big)  \Bigg]
\end{eqnarray}
is the larger positive root of
\begin{eqnarray}
    r-\frac{2}{3}\Lambda r^3-M_{\text{B}}=0  \, .
\end{eqnarray}
These bounds arise from $\rho-P\geqslant0$ and $\rho+2P\geqslant0$, respectively, and therefore from the dominant and strong energy conditions.

\begin{figure}[tbp]
    \centering
    \includegraphics[width=10cm]{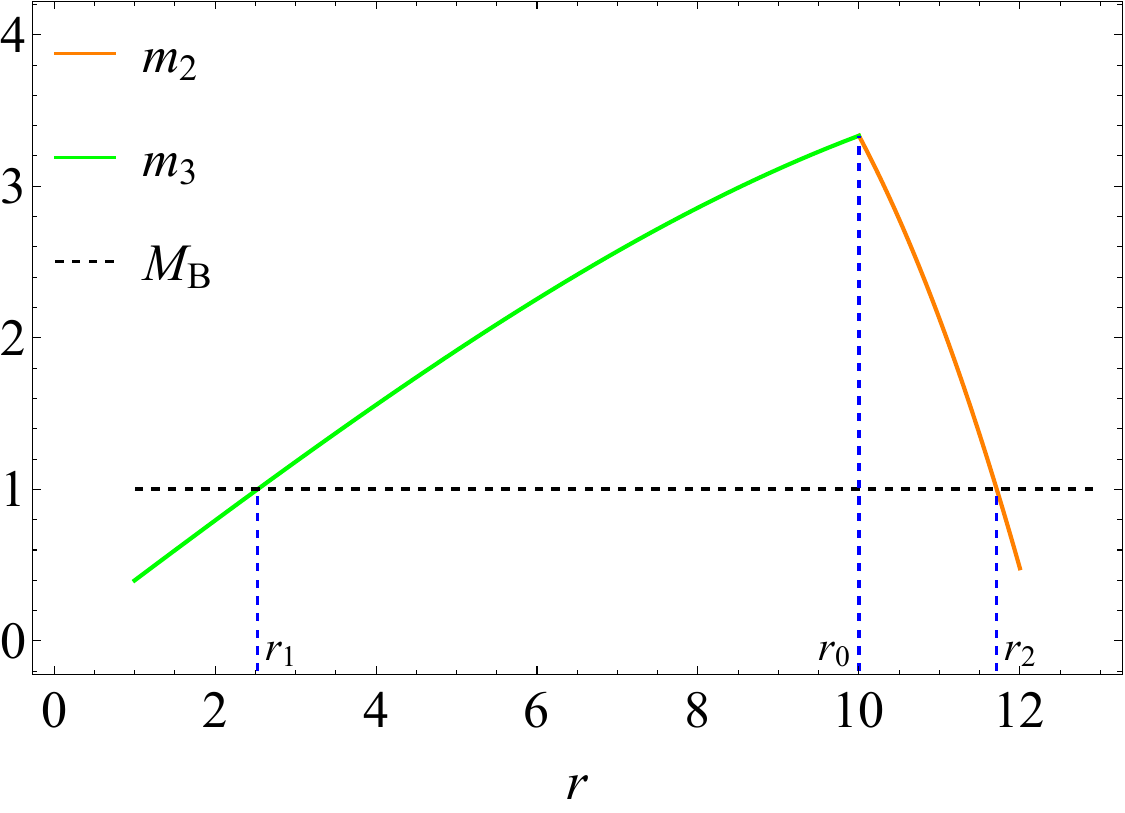}
    \caption{Allowed region for $m(r)$ at $M_{\text{B}}=1$ and $\Lambda=0.01$. Within the halo, $m(r)$ must be nondecreasing and remain between $M_{\text{B}}$ (dashed line) and the upper envelope formed by $m_3$ (green) and $m_2$ (orange).}
    \label{m23}
\end{figure}

Equation~(\ref{pE}) explains why the halo cannot extend to either nondegenerate horizon: the denominator of $P/\rho$ tends to zero there, so matter arbitrarily close to a horizon would violate the dominant energy condition $\rho\geqslant |P|$. Figure~\ref{m23} shows the allowed region in the $(r,m)$ plane. Since $m'=4\pi r^2\rho\geqslant0$, the mass function must be nondecreasing. At $r_2$, the upper bound returns to $M_{\text{B}}$, so a nonzero halo that has already accumulated mass cannot reach this endpoint. Its allowed distribution range is therefore $[r_{1}, r_{2})$.

From FIG. \ref{m23}, we can also see that the ordinate of the intersection point of $m_2(r)$ and $m_3(r)$ must be greater than $M_{\text{B}}$, i.e.,
\begin{eqnarray}
    m_2(r_0)=m_3(r_0)=\frac{1}{3\sqrt{\Lambda}}>M_{\text{B}}   \, .
\end{eqnarray}
The maximum of the upper envelope is $1/(3\sqrt{\Lambda})$. Once a nonzero halo mass has been enclosed, the mass bound is
\begin{eqnarray}
    M_{\text{B}}<m(r)<\frac{1}{3\sqrt{\Lambda}}  \, .
\end{eqnarray}
At the inner halo edge and in the inner vacuum region, $m=M_{\text{B}}$. This implies an upper bound on the mass of the black hole. It coincides with the Nariai mass of Schwarzschild--de Sitter spacetime; equality corresponds to the degenerate limit rather than a configuration with two distinct horizons.

The constraints on the matter support can be summarized as
\begin{align}\label{mLambda2}
    \rho&\geqslant0 \, ,  \qquad  M_{\text{B}}<\frac{1}{3\sqrt{\Lambda}}\, ,\qquad &r_{1}\leqslant r< r_{2}   \, ,\nonumber\\
    M_{\text{B}}&\leqslant m(r)\leqslant    \frac{2}{5}r-\frac{1}{15}\Lambda r^3\, ,\qquad &r_{1}\leqslant r\leqslant r_0   \, ,\nonumber\\
    m(r_0)&\leqslant m(r) \leqslant     r-\frac{2}{3}\Lambda r^3\, ,\qquad  &r_0<r < r_{2}  \, .
\end{align}
where $r_0=1/\sqrt{\Lambda}$ and $m(r)$ is nondecreasing. These matter bounds are not imposed on vacuum intervals where $\rho=P=0$. As $\Lambda\to0$, $r_1\to2.5M_{\text{B}}$, $r_2\to\infty$, and the mass bound disappears, recovering Ref.~\cite{Shen:2024qbb}.

\section{A Hernquist-type profile}\label{S3}
The Hernquist profile~\cite{Hernquist:1990be} is
\begin{eqnarray}\label{rhoMB0}
    \rho(r)=\frac{M_{\text{DM}}a}{2\pi r(a+r)^3}   \, ,
\end{eqnarray}
where $M_{\text{DM}}$ and $a$ are the halo mass parameter and scale parameter. Halo scale radii discussed in Ref.~\cite{Shen:2023erj} range from $0.005$ to $50\,\mathrm{kpc}$. In the presence of a black hole, the vanishing-radial-pressure construction of Ref.~\cite{Cardoso:2021wlq} uses
\begin{eqnarray}
    \rho(r)=\frac{M_{\text{DM}}(a+2M_{\text{B}})}{2\pi r(a+r)^3}\Big( 1-\frac{2M_{\text{B}}}{r} \Big)  \, .
\end{eqnarray}
This profile vanishes at the event horizon but violates the dominant energy condition nearby. Considering the energy conditions, dark matter cannot be distributed close to the horizon. It can only exist beyond $2.5M_{\text{B}}$, and the density can be taken as \cite{Shen:2024qbb}
\begin{eqnarray}\label{rho2p5MB}
    \rho(r)=\frac{M_{\text{DM}}(a+2.5M_{\text{B}})}{2\pi r(a+r)^3}\Big( 1-\frac{2.5M_{\text{B}}}{r} \Big)  \, .
\end{eqnarray}
with $r\geqslant2.5m(r)$ required throughout the halo.

For $\Lambda>0$, we introduce an outer cutoff at $C_1r_0$, with
\begin{eqnarray}
    \frac{r_{1}}{r_0}\leqslant C_1 <\frac{r_{2}}{r_0}  \, ,
\end{eqnarray}
and take
\begin{eqnarray}
    \rho(r)=\frac{M_{\text{DM}}(a+r_{1})}{2\pi r(a+r)^3}\Big( 1-\frac{r_{1}}{r} \Big)\Big( 1-\frac{r}{C_1r_0} \Big)  \, .
\end{eqnarray}
on $r_1\leqslant r\leqslant C_1r_0$, with $\rho=0$ outside this interval. The two cutoff factors make the density vanish at both boundaries. We set $C_1=1$ in the following calculations:
\begin{eqnarray}\label{rhoH}
    \rho(r)=\frac{M_{\text{DM}}(a+r_{\text{1}})}{2\pi r(a+r)^3}\Big( 1-\frac{r_{1}}{r} \Big)\Big( 1-\sqrt{\Lambda}r \Big)  \, ,\quad r_1 \leqslant r \leqslant r_0\, .
\end{eqnarray}
The parameters must additionally satisfy the pointwise mass bounds in Eq.~(\ref{mLambda2}). Integrating Eq.~(\ref{rhoE}) with $m(r_1)=M_{\text{B}}$ gives
\begin{eqnarray}\label{mH}
    m(r)&=&M_{\text{B}}+\frac{M_{\text{DM}}(r-r_{1})^2}{(a+r)^2} +\sqrt{\Lambda}\frac{M_{\text{DM}}(r-r_{1})[ (3a+2r_{1})r+2a^2+ar_{1}]}{(a+r)^2} \nonumber\\
    &&-2\sqrt{\Lambda}M_{\text{DM}}(a+r_{1})\ln\frac{a+r}{a+r_{1}}
\end{eqnarray}
for $r_1\leqslant r\leqslant r_0$. In the vacuum regions, $m=M_{\text{B}}$ for $r<r_1$ and $m=m(r_0)$ for $r>r_0$.  For $\Lambda>0$, $M_{\text{DM}}$ is a normalization parameter and need not equal this mass. However, it still reflects the dark matter halo mass, i.e., a larger $M_{\text{DM}}$ implies a more massive dark matter halo. For $\Lambda>0$, $M_{\text{DM}}$ is a normalization parameter and need not equal this mass. The limit $\Lambda\to0$ recovers Eq.~(\ref{rho2p5MB}), and subsequently setting $M_{\text{B}}=0$ recovers Eq.~(\ref{rhoMB0}).

The radial metric function is
\begin{equation}\label{g}
g_{rr}^{-1}(r)\equiv g(r) =
\begin{dcases}
    1-\frac{2M_{\text{B}}}{r}-\frac{\Lambda r^2}{3}  \, , &   r<r_{1} \, ,\\[6pt]
    1-\frac{2m(r)}{r}-\frac{\Lambda r^2}{3}  \, ,     &   r_{1}\leqslant r\leqslant r_0\, , \\[6pt]
    1-\frac{2m(r_0)}{r}-\frac{\Lambda r^2}{3}  \, ,   &   r>r_0   \, .
\end{dcases}
\end{equation}
Equation~(\ref{fE}) can be written as
\begin{eqnarray}\label{fg}
    \frac{f'(r)}{f(r)}=\frac{\Big( 1-\frac{2m(r)}{r}-\frac{\Lambda r^2}{3} \Big)'+\frac{2m'(r)}{r}}{1-\frac{2m(r)}{r}-\frac{\Lambda r^2}{3}}=\frac{g'(r)}{g(r)}+\frac{8\pi r\rho(r)}{g(r)}   \, .
\end{eqnarray}
In vacuum, $f\propto g$. We fix the time normalization by choosing\footnote{Choosing $f=g$ in the inner vacuum region $r<r_1$ instead would multiply $f$ everywhere by a constant, equivalent to a constant rescaling of $t$.}
\begin{eqnarray}\label{f1}
    f(r)=g(r)=1-\frac{2m(r_0)}{r}-\frac{\Lambda r^2}{3} \,,\qquad r\geqslant r_0 \, .
\end{eqnarray}
Integrating inward through the halo yields
\begin{eqnarray}\label{f2}
    f(r)&=&\Big( 1-\frac{2m(r_0)}{r_0}-\frac{\Lambda r_0^2}{3} \Big)\exp\Bigg[\int_{r_0}^r\frac{2}{r^2}\frac{m(r)-\frac{\Lambda r^3}{3}}{1-\frac{2m(r)}{r}-\frac{\Lambda r^2}{3}}dr \Bigg]     \,,\qquad r_{1}<r< r_{0}\, .
\end{eqnarray}
Continuity at $r_1$ then gives
\begin{eqnarray}\label{f3}
    f(r)=C_2 g(r)=C_2 \Big(1-\frac{2M_{\text{B}}}{r}-\frac{\Lambda r^2}{3} \Big) \,, \qquad r< r_{1} \, ,
\end{eqnarray}
where
\begin{eqnarray}\label{Constant_C2}
    C_2 = \frac{ 1-\frac{2m(r_0)}{r_0}-\frac{\Lambda r_0^2}{3}}{1-\frac{2M_{\text{B}}}{r_{1}}-\frac{\Lambda r_{1}^2}{3}} \exp\Bigg[\int_{r_0}^{r_{1}}\frac{2}{r^2}\frac{m(r)-\frac{\Lambda r^3}{3}}{1-\frac{2m(r)}{r}-\frac{\Lambda r^2}{3}}\mathrm{d}r \Bigg]   \, .
\end{eqnarray}
The density is continuous and vanishes at $r_1$ and $r_0$, so $m$ and $m'$ are continuous across both interfaces. Equations~(\ref{g}) and (\ref{fg}), together with the matching constants, make $f$, $g$, and their first derivatives continuous. The interfaces therefore carry no thin shells. Higher derivatives need not be continuous, a point relevant to the numerical treatment in Sec.~\ref{S5}.

\section{Black hole images in a Hernquist-type dark matter halo}\label{S4}
We use ray tracing to compare the optical appearance of the halo spacetimes with and without a positive cosmological constant. The images depend on the photon effective potential and the emission model. We adopt the thin-disk emission profile
\begin{eqnarray}
    I_{\text{em}}(r) = \left\{
    \begin{array}{ll}
    \frac{1}{(r-r_{\text{h}}+1)^3}  \, ,  \qquad&   r\geqslant r_{\text{h}} \, , \\
    0  \,, \qquad&  r<r_{\text{h}}  \, ,
    \end{array}
 \right.
\end{eqnarray}
where $r_{\text{h}}$ is the event-horizon radius. Lengths in this prescription and in the numerical plots are expressed in units of $M_{\text{B}}$. For $\Lambda>0$, the observer is placed at $r_{\text{obs}}=0.8r_{\text{c}}$. In the case $\Lambda=0$, we set $r_{\text{obs}}=300$, since at such a large distance the observed image differs negligibly from the one seen at infinity.

With impact parameter $b$, the radial null geodesic equation is
\begin{eqnarray}
    \frac{f(r)}{g(r)}\dot{r}^2+V_{\text{eff}}(r)=\frac{1}{b^2}   \, ,
\end{eqnarray}
where
\begin{eqnarray}
    V_{\text{eff}}(r)=\frac{f(r)}{r^2}
\end{eqnarray}
is the photon effective potential. Here $b=L/E$, with $E$ and $L$ the conserved photon energy and angular momentum. The overdot denotes a derivative with respect to a rescaled affine parameter. Local maxima of $V_{\text{eff}}$ locate unstable circular photon orbits.

A useful family of halo profiles is~\cite{Dehnen:1993uh,Hernquist:1990be,Shen:2023erj,Graham:2005xx,Figueiredo:2023gas,Bolokhov:2025zva}
\begin{eqnarray}
    \rho(r)=2^{(\gamma-\alpha)/k}\rho_0(r/a)^{-\alpha}\Big[ 1+ \Big( \frac{r}{a} \Big)^k\Big]^{-(\gamma-\alpha)/k}   \, ,
\end{eqnarray}
where $\rho_0$ is the density at $r=a$ before the cutoff factors are introduced. We generalize the construction of Sec.~\ref{S3} by taking
\begin{eqnarray}
    \rho(r)=2^{(\gamma-\alpha)/k}\rho_0(r/a)^{-\alpha}\Big[1+ \Big( \frac{r}{a} \Big)^k\Big]^{-(\gamma-\alpha)/k}\left(1-\frac{r_{1}}{r}\right)\left(1-\sqrt{\Lambda}r\right)   \, .
\end{eqnarray}
on $r_1\leqslant r\leqslant r_0$, and zero density outside the halo. The corresponding mass function follows from
\begin{eqnarray}
    m(r)=M_{\text{B}}+\int_{r_1}^r 4\pi x^2 \rho(x) \mathrm{d}x   \, .
\end{eqnarray}
The metric is obtained by the same integration and matching procedure as in Sec.~\ref{S3}. For $\Lambda=0$, the inner edge is $2.5M_{\text{B}}$ and the outer cutoff is removed. Each choice of profile parameters must satisfy Eq.~(\ref{mLambda2}) within its matter support.

\begin{figure}[tbp]
		\begin{minipage}[t]{0.45\linewidth}
			\centering
			\includegraphics[height=5cm]{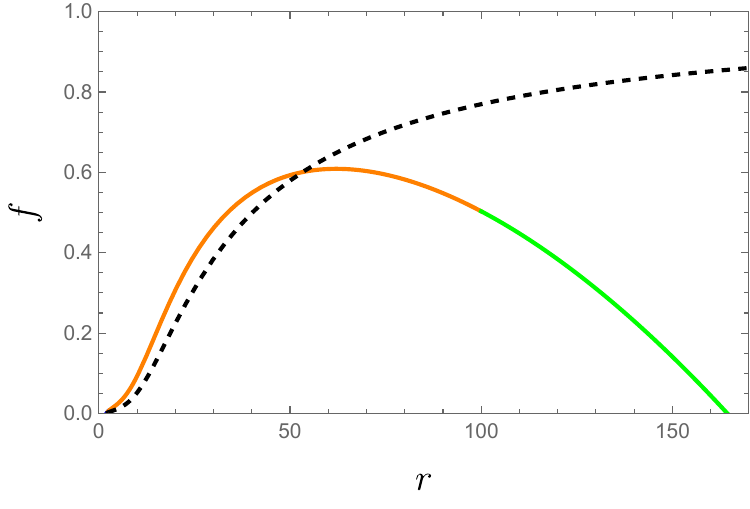}
        \par\vspace{1mm}
		\end{minipage}%
        \hspace{5mm}
		\begin{minipage}[t]{0.45\linewidth}
			\centering
			\includegraphics[height=5cm]{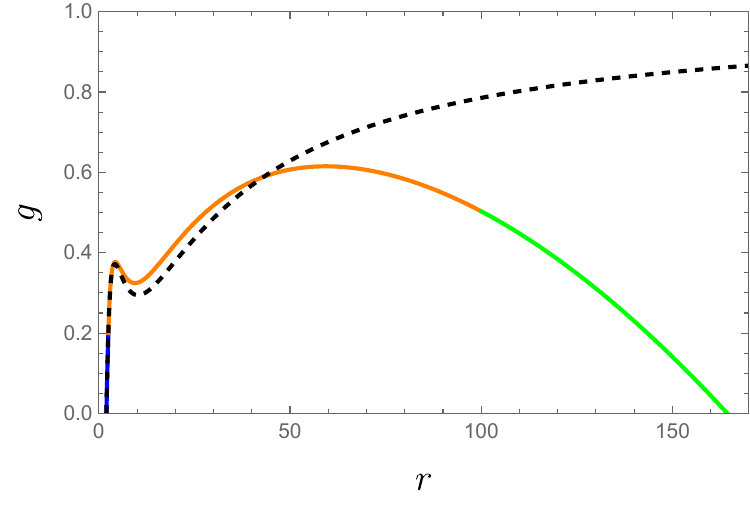}
        \par\vspace{1mm}
		\end{minipage}%
  \caption{Hernquist-type metric functions $f(r)$ (left) and $g(r)$ (right) for $a=6M_{\text{B}}$ and $M_{\text{DM}}=11.5M_{\text{B}}$. Colored curves correspond to $\Lambda=10^{-4}M_{\text{B}}^{-2}$, with blue, orange, and green indicating $r<r_1$, $r_1\leqslant r\leqslant r_0$, and $r>r_0$, respectively. Black curves show $\Lambda=0$.}
  \label{Hernfg}
\end{figure}

For $(\alpha,\gamma,k)=(1,4,1)$, the density and mass function are given by Eqs.~(\ref{rhoH}) and (\ref{mH}), with the corresponding normalization of $\rho_0$. We use $a=6M_{\text{B}}$, $M_{\text{DM}}=11.5M_{\text{B}}$, and $\Lambda=10^{-4}M_{\text{B}}^{-2}$. Figure~\ref{Hernfg} compares $f(r)$ and $g(r)$ with their $\Lambda=0$ counterparts. The differences are modest at small radii and become more pronounced farther out.

\begin{figure}[tbp]
		\begin{minipage}[t]{0.45\linewidth}
			\centering
			\includegraphics[height=4cm]{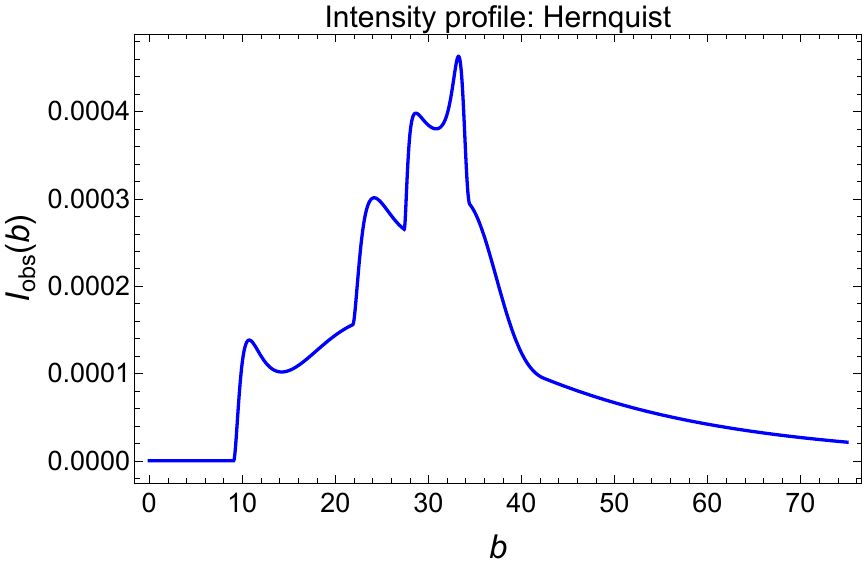}
        \par\vspace{1mm}
		\end{minipage}%
		\begin{minipage}[t]{0.45\linewidth}
			\centering
			\includegraphics[height=4cm]{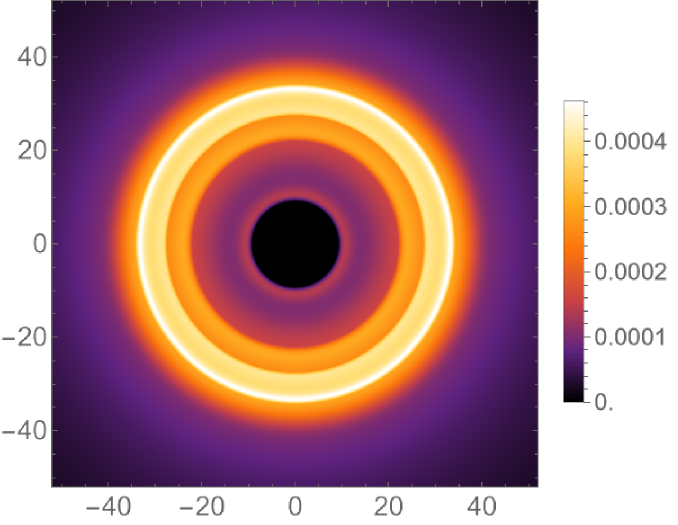}
        \par\vspace{1mm}
		\end{minipage}%

		\begin{minipage}[t]{0.45\linewidth}
			\centering
			\includegraphics[height=4cm]{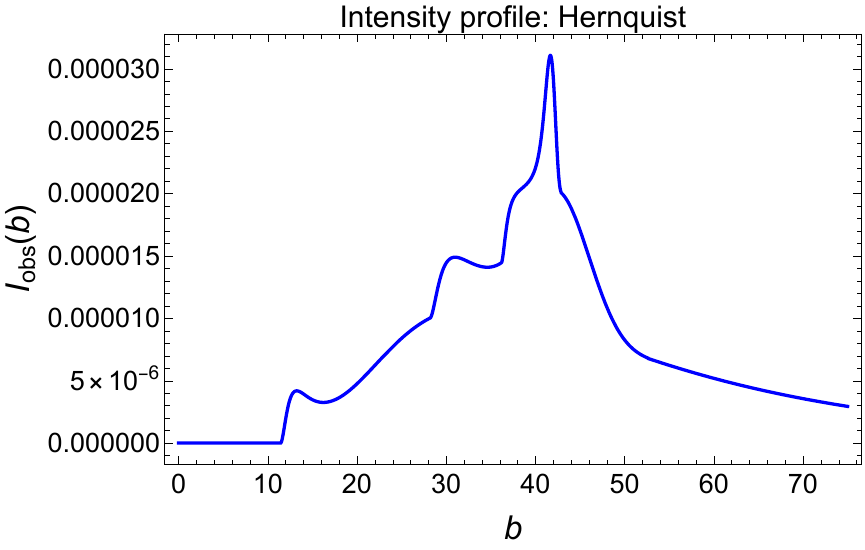}
        \par\vspace{1mm}
		\end{minipage}%
		\begin{minipage}[t]{0.45\linewidth}
			\centering
			\includegraphics[height=4cm]{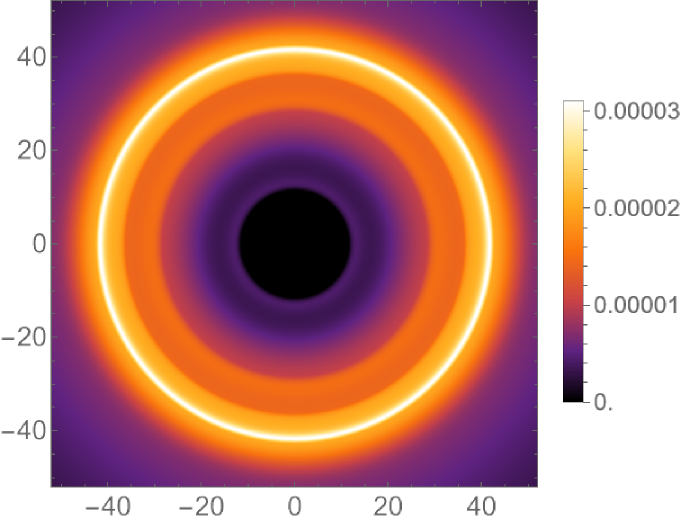}
        \par\vspace{1mm}
		\end{minipage}%
  \caption{Observed intensity $I_{\text{obs}}(b)$ (left) and images (right) for the Hernquist-type halo with $a=6M_{\text{B}}$ and $M_{\text{DM}}=11.5M_{\text{B}}$. The upper and lower rows correspond to $\Lambda=10^{-4}M_{\text{B}}^{-2}$ and $\Lambda=0$, respectively.}
  \label{Hern}
\end{figure}

For many parameter choices, the images retain a similar morphology, with changes mainly in the shadow and ring sizes. In the chosen observer setup, the positive-$\Lambda$ images are brighter, as also found in Ref.~\cite{Cao:2024kht}. The example in FIG.~\ref{Hern} shows a more substantial change: a broad bright band replaces the single prominent ring of the $\Lambda=0$ image.

\begin{figure}[tbp]
		\begin{minipage}[t]{0.45\linewidth}
			\centering
			\includegraphics[height=5cm]{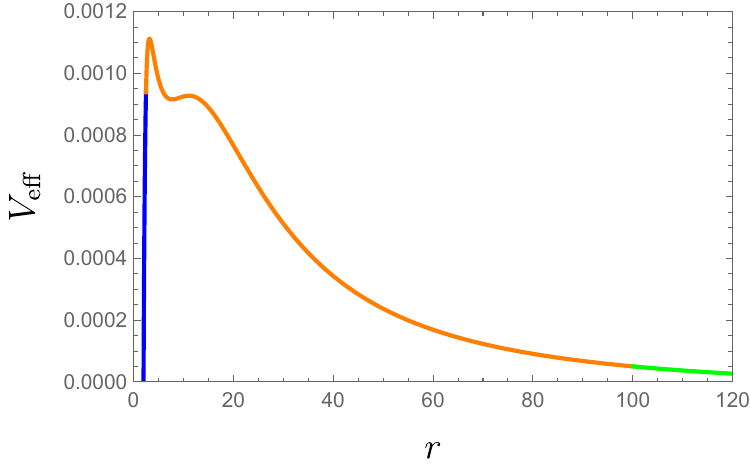}
        \par\vspace{1mm}
		\end{minipage}%
        \hspace{5mm}
		\begin{minipage}[t]{0.45\linewidth}
			\centering
			\includegraphics[height=5cm]{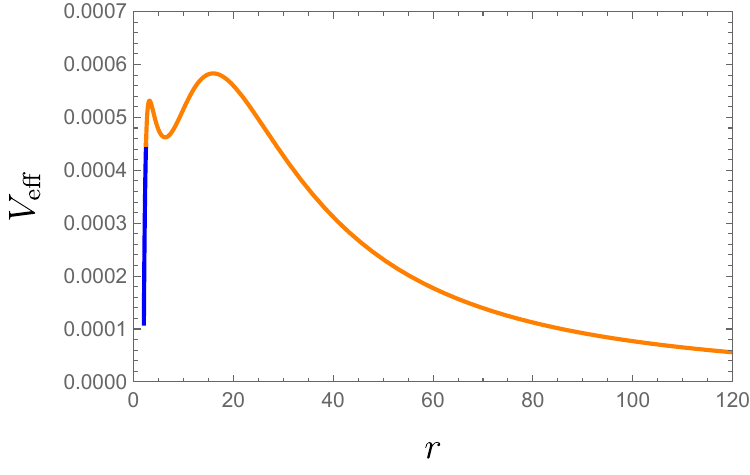}
        \par\vspace{1mm}
		\end{minipage}%
  \caption{Photon effective potentials for the Hernquist-type example: $\Lambda=10^{-4}M_{\text{B}}^{-2}$ (left) and $\Lambda=0$ (right). In the left panel, blue, orange, and green denote the inner vacuum, halo, and outer vacuum regions. In the right panel, blue and orange denote $r<2.5M_{\text{B}}$ and $r\geqslant2.5M_{\text{B}}$. The remaining parameters are those of FIG.~\ref{Hernfg}.}
  \label{HernV}
\end{figure}

For $\Lambda>0$, the bright band lies near $b\simeq29$--$33M_{\text{B}}$, whereas the prominent ring for $\Lambda=0$ lies near $b\simeq42M_{\text{B}}$. Figure~\ref{HernV} relates this difference to the potential barriers. In the positive-$\Lambda$ example, two photon spheres contribute to the band. For $\Lambda=0$, the outer potential maximum is higher than the inner one, so only the outer ring is observable in the image.

The brightness difference is associated with the redshift factor for the chosen observer near the cosmological horizon. The morphology reflects both the potential structure and the radial distribution of the emitting material. In this construction, changing $\Lambda$ also changes the halo cutoff and density. The comparison is therefore not a variation of the propagation geometry alone. Changes in $M_{\text{DM}}$ or $a$ can likewise alter the number or relative heights of the potential peaks and produce similar image features.

We also analyze black hole images in several other dark matter halo models and compare the differences between the cases with and without $\Lambda$. Details are given in the Appendix~\ref{S7}.

\section{Quasinormal modes}\label{S5}
In this section, we investigate the scalar quasinormal modes of the Hernquist-type spacetime constructed in Sec. \ref{S3}.
For the $ \Lambda > 0$ configurations considered below, the cosmological horizon $ r_\text{c} > r_0$ lies in the outer vacuum region and satisfies
\begin{eqnarray}
    1-\frac{2m(r_0)}{r_{\text{c}}}-\frac{\Lambda r_{\text{c}}^2}{3}=0\, .
\end{eqnarray}
We consider a massless scalar field satisfying the Klein--Gordon equation $\square \Phi=0$. Define $\Phi(t,r,\theta,\phi)=\mathrm{e}^{-\mathrm{i}\omega t}Y_{\ell m}(\theta,\phi)\Psi(r)/r$, we have
\begin{eqnarray}\label{Psi_equation}
    \Bigg\{fg\frac{\mathrm{d}^2}{\mathrm{d}r^2}+\frac{(fg)^{\prime}}{2}\frac{\mathrm{d}}{\mathrm{d}r}+\Big[\omega^2-\frac{f \ell(\ell+1)}{r^2}-\frac{(fg)^{\prime}}{2r}\Big]\Bigg\}\Psi(r)=0\, ,
\end{eqnarray}
where prime denotes the derivative of $r$, and the functions $f(r)$ and $g(r)$ are given by Eq. (\ref{g}), Eq. (\ref{f1}), Eq. (\ref{f2}) and Eq. (\ref{f3}).
Define the tortoise coordinate $\mathrm{d}x/\mathrm{d}r=1/\sqrt{fg}$, one gets the following Schr\"{o}dinger-like equation
\begin{eqnarray}
    \Big[\frac{\mathrm{d}^2}{\mathrm{d}x^2}+\Big(\omega^2-V(r)\Big)\Big]\Psi(r)=0\, ,
\end{eqnarray}
with the scalar field potential $V(r)$ given by
\begin{eqnarray}\label{effective_potential_QNMs}
    V(r)=f(r)\frac{\ell(\ell+1)}{r^2}+\frac{(f(r)g(r))^{\prime}}{2r}\, .
\end{eqnarray}
It is easy to find that $r_{\text{h}}$ and $r_{\text{c}}$ are regular singular points of the above equation (\ref{Psi_equation}). Therefore, according to the QNM boundary conditions, the ansatz of QNM solution $\Psi$ is
\begin{eqnarray}\label{ansatz}
    \Psi(r)=(r-r_{\text{h}})^{-\frac{\mathrm{i}\omega}{\sqrt{C_2}g^{\prime}(r_{\text{h}})}}(r-r_{\text{c}})^{\frac{\mathrm{i}\omega}{g^{\prime}(r_{\text{c}})}}u(r)\, ,
\end{eqnarray}
where $u(r)$ is a regular function on $[r_{\text{h}},r_{\text{c}}]$, and $C_2>0$ is given by Eq. (\ref{Constant_C2}). In addition, two quantities $g^{\prime}(r_{\text{h}})$ and $g^{\prime}(r_{\text{c}})$ in Eq. (\ref{ansatz}) are
\begin{eqnarray}
    g^{\prime}(r_{\text{h}})=\frac{2M_{\text{B}}}{r_{\text{h}}^2}-\frac{2\Lambda r_{\text{h}}}{3}\, ,\quad \text{and} \quad g^{\prime}(r_{\text{c}})=\frac{2m(r_0)}{r_{\text{c}}^2}-\frac{2\Lambda r_{\text{c}}}{3}\, .
\end{eqnarray}
Considering that $f(r)$ and $g(r)$ are merely functions with continuous first derivatives, from Eq. (\ref{effective_potential_QNMs}), we know that the effective potential $V(r)$ is merely a continuous function on $[r_\text{h},r_\text{c}]$. Therefore, in order to minimize numerical errors, we adopt the multi-domain pseudospectral method~\cite{PanossoMacedo:2024pox,Xie:2025jbr}. For convenience, we denote that the solutions on the intervals $[r_{\text{h}},r_{\text{1}}]$, $[r_{\text{1}},r_0]$ and $[r_0,r_{\text{c}}]$ are $u_{(1)}(r)$, $u_{(2)}(r)$ and $u_{(3)}(r)$, respectively. Substituting Eq. (\ref{ansatz}) into Eq. (\ref{Psi_equation}), one obtains the equations that $u_1(r)$, $u_2(r)$ and $u_3(r)$ satisfy on their respective intervals. For $k=1,2,3$, the expressions can be written as follows
\begin{eqnarray}\label{numerical_Eq}
    &&\Bigg\{{}^{[0]}C_2^{(k)}(r)\frac{\mathrm{d}^2}{\mathrm{d}r^2}+\Big[{}^{[0]}C_1^{(k)}(r)+{}^{[1]}C_1^{(k)}(r)\omega\Big]\frac{\mathrm{d}}{\mathrm{d}r}\nonumber\\
    &&+\Big[{}^{[0]}C_0^{(k)}(r)+{}^{[1]}C_0^{(k)}(r)\omega+{}^{[2]}C_0^{(k)}(r)\omega^2\Big]\Bigg\}u_{(k)}(r)=0\, ,
\end{eqnarray}
where the above coefficients ${}^{[0]}C_2^{(k)}(r)$, ${}^{[0]}C_1^{(k)}(r)$, ${}^{[1]}C_1^{(k)}(r)$, ${}^{[0]}C_0^{(k)}(r)$, ${}^{[1]}C_0^{(k)}(r)$ and ${}^{[2]}C_0^{(k)}(r)$ with $k=1,2,3$ can be obtained easily, although their expressions are a bit long. In terms of more computational details, at the connection surfaces $r=r_1$ and $r=r_0$, we require that the solutions and their derivatives are continuous. This means that for $r=r_1$, one gets
\begin{eqnarray}\label{condition_r1}
    u_{(1)}|_{r=r_1}=u_{(2)}|_{r=r_1}\, ,\quad \text{and}\quad \frac{\mathrm{d}u_{(1)}}{\mathrm{d}r}\Bigg|_{r=r_1}=\frac{\mathrm{d}u_{(2)}}{\mathrm{d}r}\Bigg|_{r=r_1}\, ,
\end{eqnarray}
and for $r=r_0$, one gets
\begin{eqnarray}\label{condition_r0}
    u_{(2)}|_{r=r_0}=u_{(3)}|_{r=r_0}\, ,\quad \text{and}\quad \frac{\mathrm{d}u_{(2)}}{\mathrm{d}r}\Bigg|_{r=r_0}=\frac{\mathrm{d}u_{(3)}}{\mathrm{d}r}\Bigg|_{r=r_0}\, .
\end{eqnarray}
After discretization, four conditions (\ref{condition_r1}) and (\ref{condition_r0}) can be written into four linear equations associated with the state vector $\mathbf{u}$ which is given by
\begin{eqnarray}
    \mathbf{u}=\Big(\mathbf{u}_{(1)},\mathbf{u}_{(2)},\mathbf{u}_{(3)}\Big)^{\mathrm{T}}=\Big(u_{(1)}^0,\cdots,u_{(1)}^{N_1},u_{(2)}^0,\cdots,u_{(2)}^{N_2},u_{(3)}^0,\cdots,u_{(3)}^{N_3}\Big)^{\mathrm{T}}\, .
\end{eqnarray}
Combine with the bulk discretization results of Eq. (\ref{numerical_Eq}) and the interface discretization results from Eqs. (\ref{condition_r1}) and Eqs. (\ref{condition_r0}), one obtains that
\begin{eqnarray}\label{T_omega_problem}
    \mathbf{T}(\omega)\mathbf{u}=0\, .
\end{eqnarray}
Following the approach used in~\cite{Jansen:2017oag,Li:2024npg,Li:2025ljb,Xie:2025jbr,Yu:2026rku,Xie:2026yjg}, the polynomial eigenvalue problem for the matrix $\mathbf{T}(\omega)$ can be transformed into a generalized eigenvalue problem. The obtained QNM spectra are checked for convergence with increasing resolution.

\begin{figure}[htbp]
    \centering
    \begin{minipage}[b]{0.33\linewidth}
        \centering
        {\footnotesize\makebox[\linewidth][l]{\hspace{0.14\linewidth}\textbf{(a)}}\par}\vspace{2pt}
        \includegraphics[width=\linewidth]{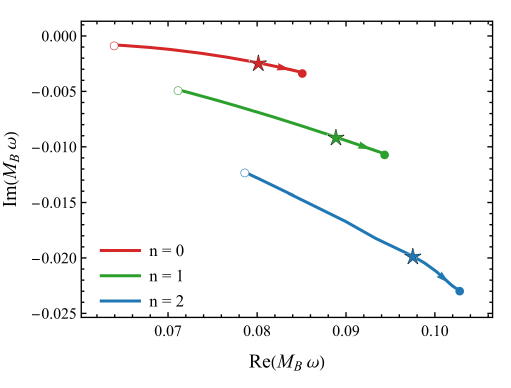}
    \end{minipage}%
    \hfill
    \begin{minipage}[b]{0.33\linewidth}
        \centering
        {\footnotesize\makebox[\linewidth][l]{\hspace{0.14\linewidth}\textbf{(b)}}\par}\vspace{2pt}
        \includegraphics[width=\linewidth]{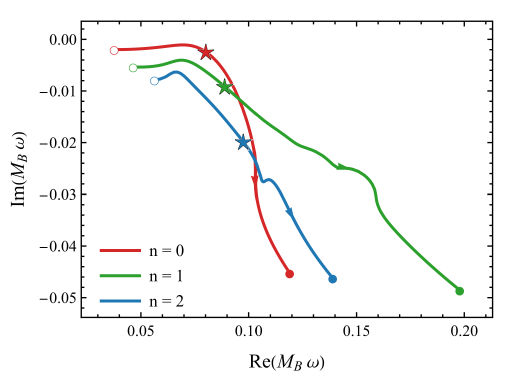}
    \end{minipage}%
    \hfill
    \begin{minipage}[b]{0.33\linewidth}
        \centering
        {\footnotesize\makebox[\linewidth][l]{\hspace{0.14\linewidth}\textbf{(c)}}\par}\vspace{2pt}
        \includegraphics[width=\linewidth]{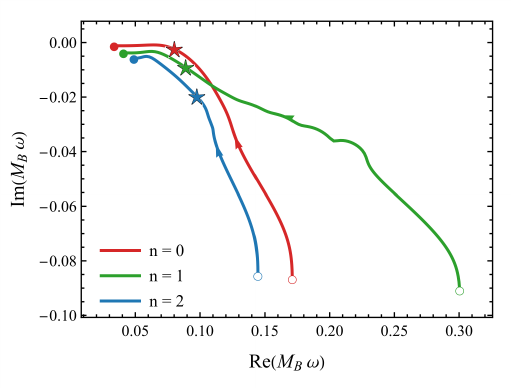}
    \end{minipage}
    \caption{Trajectories of three selected $\ell=2$ scalar QNMs in the complex-frequency plane. Red, green, and blue denote the modes labeled $n=0,1,2$, respectively, at the common reference point $(\Lambda M_{\text{B}}^2,a/M_{\text{B}},M_{\text{DM}}/M_{\text{B}})=(10^{-4},6,11.5)$. In panel (a), $\Lambda M_{\text{B}}^2$ varies from $7\times10^{-6}$ to $1.5\times10^{-4}$, with $a/M_{\text{B}}=6$ and $M_{\text{DM}}/M_{\text{B}}=11.5$ fixed. In panel (b), $a/M_{\text{B}}$ varies from $4.5$ to $20$, with $\Lambda M_{\text{B}}^2=10^{-4}$ and $M_{\text{DM}}/M_{\text{B}}=11.5$ fixed. In panel (c), $M_{\text{DM}}/M_{\text{B}}$ varies from $1/16$ to $14$, with $\Lambda M_{\text{B}}^2=10^{-4}$ and $a/M_{\text{B}}=6$ fixed. Open and filled circles mark the beginning and end of each trajectory, respectively. Stars mark the common reference point. The arrows indicate the direction of increasing parameter.
    }
    \label{QNMmigration}
\end{figure}

We first examine how the scalar QNMs change as the model parameters are varied. We use $M_{\text{B}}$ as the reference mass scale and present the QNM results in terms of the dimensionless quantities $M_{\text{B}}\omega$, $\Lambda M_{\text{B}}^2$, $a/M_{\text{B}}$, and $M_{\text{DM}}/M_{\text{B}}$. FIG. \ref{QNMmigration} shows the trajectories of three selected $\ell=2$ modes in the complex-frequency plane. The modes are labeled $n=0,1,2$ at the common reference point $(\Lambda M_{\text{B}}^2,a/M_{\text{B}},M_{\text{DM}}/M_{\text{B}})=(10^{-4},6,11.5)$ and are followed continuously as one parameter is varied.

As shown in FIG. \ref{QNMmigration}(a), increasing $\Lambda$ moves all three modes toward larger $\operatorname{Re}(M_{\text{B}}\omega)$ and more negative $\operatorname{Im}(M_{\text{B}}\omega)$. In FIG. \ref{QNMmigration}(b), increasing $a$ moves the modes overall toward larger $\operatorname{Re}(M_{\text{B}}\omega)$, while their imaginary parts first approach the real axis and then become more negative. The $n=1$ mode extends farther in the real-frequency direction, whereas the $n=0$ and $n=2$ trajectories show more pronounced turning. In FIG. \ref{QNMmigration}(c), increasing $M_{\text{DM}}$ moves the three modes toward smaller $\operatorname{Re}(M_{\text{B}}\omega)$ and less negative $\operatorname{Im}(M_{\text{B}}\omega)$. In addition, overtakings of overtones are also happened in FIG. \ref{QNMmigration}(b) and FIG. \ref{QNMmigration}(c).

FIG. \ref{QNMmigration} illustrates the parameter dependence of several selected modes. We can not only track how a given mode moves as the parameters vary, but also obtain the distribution of QNMs for a given set of parameters. We next examine the distribution of the QNM spectra for different values of $\Lambda$, together with the corresponding effective potentials. It is worth mentioning that, for $\Lambda=0$, spatial infinity replaces the cosmological horizon as the outer boundary, and the corresponding ansatz like Eq. (\ref{ansatz}) should be changed. Here, we omitted this ansatz explicitly. We use the same multi-interval pseudospectral approach after compactifying the exterior region. The resulting quadratic eigenvalue problem is again transformed into a generalized eigenvalue problem, and the convergent eigenvalues are identified as the QNM spectra.

\begin{figure}[htbp]
    \centering
    \begin{minipage}[b]{0.325\linewidth}
        \centering
        {\footnotesize\makebox[\linewidth][l]{\hspace{0.25\linewidth}\textbf{(a)}}\par}\vspace{2pt}
        \includegraphics[width=\linewidth]{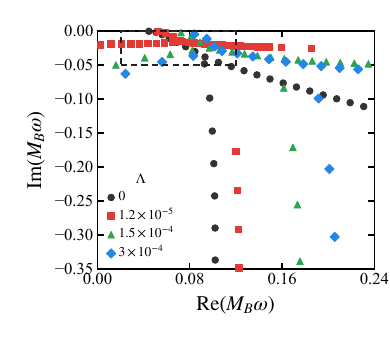}
    \end{minipage}%
    \hfill
    \begin{minipage}[b]{0.325\linewidth}
        \centering
        {\footnotesize\makebox[\linewidth][l]{\hspace{0.25\linewidth}\textbf{(b)}}\par}\vspace{2pt}
        \includegraphics[width=\linewidth]{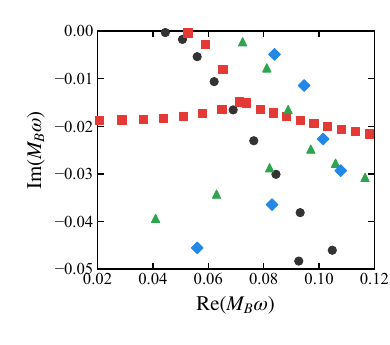}
    \end{minipage}%
    \hfill
    \begin{minipage}[b]{0.325\linewidth}
        \centering
        {\footnotesize\makebox[\linewidth][l]{\hspace{0.25\linewidth}\textbf{(c)}}\par}\vspace{2pt}
        \includegraphics[width=\linewidth]{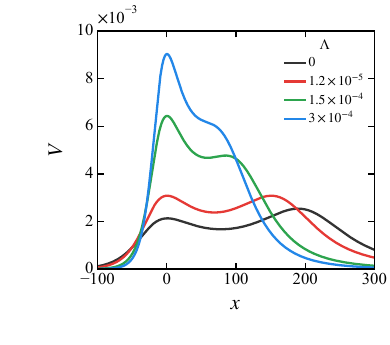}
    \end{minipage}
    \caption{Panel (a): Scalar QNM spectrum distributions for $a=7.5M_{\text{B}}$, $M_{\text{DM}}=14M_{\text{B}}$, and $\ell=2$. Black points, red filled squares, green filled triangles, and blue filled diamonds correspond to $\Lambda M_{\text{B}}^2=0$, $1.2\times10^{-5}$, $1.5\times10^{-4}$, and $3\times10^{-4}$, respectively. Panel (b): Enlargement of the boxed region in panel (a), with $0.02\leq\mathrm{Re}(M_{\text{B}}\omega)\leq0.12$ and $-0.05\leq\mathrm{Im}(M_{\text{B}}\omega)\leq0$. Panel (c): The effective potentials (\ref{effective_potential_QNMs}) for the same parameters and with the same color coding. The potential $V$ is plotted against the dimensionless tortoise coordinate $x$.}
    \label{qnm-spectrum_and_scalar-effective-potential}
\end{figure}

As shown in the left panel of FIG. \ref{qnm-spectrum_and_scalar-effective-potential}, the QNMs can be divided into two families. One family is approximately parallel to the imaginary axis, while the other is approximately parallel to the real axis. As $\Lambda$ increases, the first family shifts as a whole toward larger real frequencies. The second family becomes progressively more horizontal, with smaller imaginary parts. The corresponding effective potentials are shown in the right panel of FIG. \ref{qnm-spectrum_and_scalar-effective-potential}. For $\Lambda M_{\text{B}}^2=1.2\times10^{-5}$, the potential has a pronounced double-peak structure. As $\Lambda M_{\text{B}}^2$ increases, the second peak becomes progressively less prominent. At $\Lambda M_{\text{B}}^2=3\times10^{-4}$, the potential consists of one dominant peak followed by a broad shoulder.

The results above show how the QNM spectra changes with the model parameters. Pezzella \textit{et al.} found that, for sufficiently small $M_{\text{DM}}/a$, the relative QNM spectrum shift is approximately linear in $M_{\text{DM}}/a$~\cite{Pezzella:2024tkf}. Motivated by such result, we examine whether the same behavior appears in our model and how the corresponding coefficient depends on $\Lambda$.

\begin{figure}[htbp]
    \centering
    \includegraphics[width=10cm]{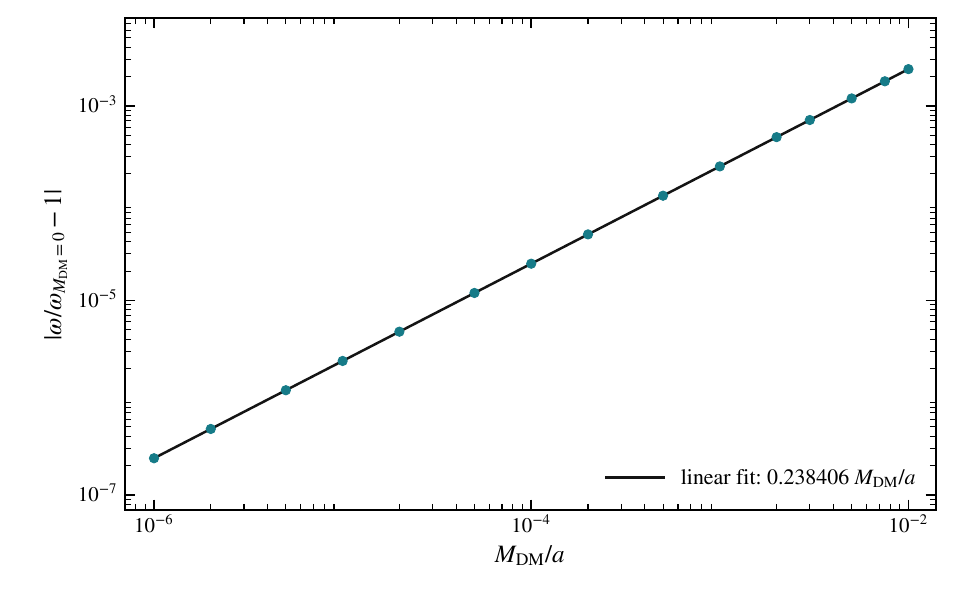}
    \caption{Absolute relative shift of the $\ell=2$, $n=0$ mode as a function of $M_{\text{DM}}/a$, for $a=20M_{\text{B}}$ and $\Lambda M_{\text{B}}^2=4\times10^{-3}$. The reference spectrum $\omega_{M_{\text{DM}}=0}$ is the corresponding Schwarzschild--de Sitter QNM spectra  at the same $\Lambda$. Circles show the numerical results, and the solid line shows the linear fit in Eq.(\ref{qnm-weak-halo-scaling}) with $C=0.238406$. Both axes are logarithmic.}
    \label{qnm-weak-halo-modulus}
\end{figure}

We take $a=20M_{\text{B}}$ and follow the $\ell=2$, $n=0$ mode continuously from the corresponding $M_{\text{DM}}=0$ solution. When $M_{\text{DM}}=0$, the dark matter contribution vanishes and the spacetime reduces to Schwarzschild--de Sitter spacetime. For each value of $\Lambda$, the spectrum is compared with $\omega_{M_{\text{DM}}=0}$ evaluated at the same $\Lambda$. For $5\times10^{-4}\leqslant\Lambda M_{\text{B}}^2\leqslant10^{-2}$ and $10^{-6}\leqslant M_{\text{DM}}/a\leqslant10^{-2}$, the numerical results are well described by
\begin{eqnarray}\label{qnm-weak-halo-scaling}
    \left|\frac{\omega}{\omega_{M_{\text{DM}}=0}}-1\right|\simeq C(\Lambda M_{\text{B}}^2) \times \frac{M_{\text{DM}}}{a}\, ,
\end{eqnarray}
where the coefficient $C(\Lambda M_{\text{B}}^2)$ is the function of $\Lambda M_{\text{B}}^2$. For $\Lambda M_{\text{B}}^2=4\times10^{-3}$, the result is shown in FIG. \ref{qnm-weak-halo-modulus}, for which the fit gives $C=0.238406$.

\begin{figure}[H]
    \centering
    \includegraphics[width=10cm]{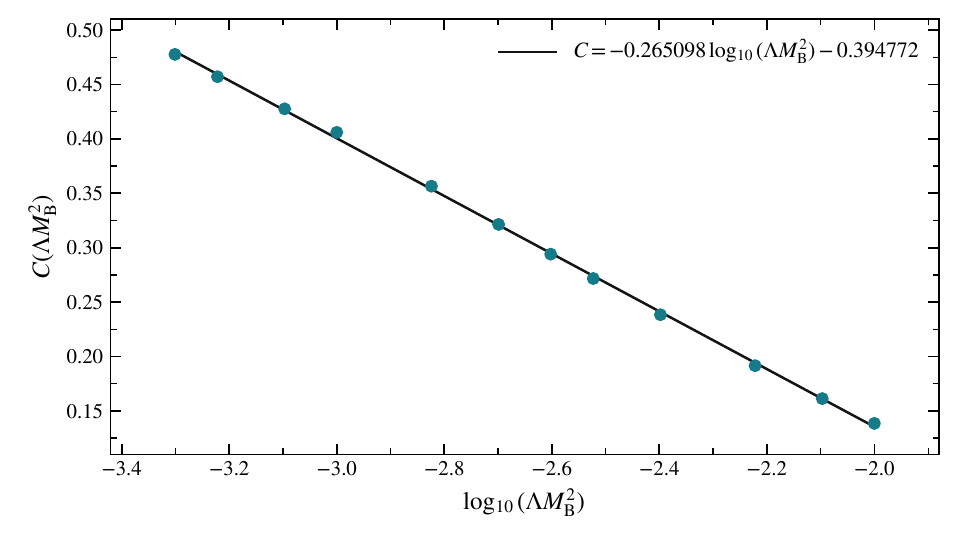}
    \caption{Dependence of the coefficient $C$ in Eq. (\ref{qnm-weak-halo-scaling}) on $\log_{10}(\Lambda M_{\text{B}}^2)$, for $a=20M_{\text{B}}$ and the $\ell=2$, $n=0$ scalar mode. Circles show the coefficients obtained for twelve values of $\Lambda M_{\text{B}}^2$ between $5\times10^{-4}$ and $10^{-2}$. The solid line shows the fit in Eq. (\ref{qnm-weak-halo-coefficient-fit}) over the same interval.}
    \label{qnm-weak-halo-coefficient}
\end{figure}

To determine how the coefficient changes with the cosmological constant, we repeat the calculation for different values of $\Lambda$, varying $M_{\text{DM}}/a$ over the same range for each case. As shown in FIG. \ref{qnm-weak-halo-coefficient}, $C$ decreases from approximately $0.48$ to $0.14$ as $\Lambda M_{\text{B}}^2$ increases from $5\times10^{-4}$ to $10^{-2}$. Over this interval, the numerical values are well represented by
\begin{eqnarray}\label{qnm-weak-halo-coefficient-fit}
    C(\Lambda M_{\text{B}}^2)\simeq-0.265098\log_{10}(\Lambda M_{\text{B}}^2)-0.394772\, .
\end{eqnarray}
This expression is used only to summarize the numerical dependence of $C$ on $\Lambda M_{\text{B}}^2$ over the parameter range considered here.

\section{Conclusion and discussion}\label{S6}
We have studied static, spherically symmetric black holes with dark matter halos, a positive cosmological constant, and vanishing radial pressure. Imposing the null, weak, strong, and dominant energy conditions on the halo gives the mass-function bounds in Eq. (\ref{mLambda2}). The halo must occupy a finite interval contained in $[r_1,r_2)$ and remain separated from both horizons. The total mass is bounded above by the Nariai mass $1/(3\sqrt{\Lambda})$.

We constructed a Hernquist-type profile supported on $[r_1,r_0]$, with $r_0=1/\sqrt{\Lambda}$, and obtained its mass function and matched metric. For admissible parameters, the density vanishes continuously at both interfaces and no thin shells are present. The $\Lambda\to0$ limit recovers the model discussed in Ref.~\cite{Shen:2024qbb}. Formally replacing the inner cutoff by $2M_{\text{B}}$ in the $\Lambda=0$ profile recovers Ref. \cite{Cardoso:2021wlq}, but that replacement does not preserve the energy-condition constraint.

The image comparison was extended to the NFW, Jaffe, Burkert, and Taylor--Silk models. In the selected configurations, positive $\Lambda$ produces brighter images and generally smaller ring scales under the adopted observer prescriptions. The Hernquist, NFW, and Burkert examples show substantial morphological changes, whereas the Jaffe images remain similar. The Taylor--Silk example illustrates that a bright band need not imply two photon spheres.
These differences depend on the effective potential, and parameters other than
$\Lambda$ can produce similar effects on the image. Therefore, they do not establish a unique or directly measurable signature of
$\Lambda$.

For the Hernquist-type spacetime, the selected scalar QNMs respond differently to changes in $\Lambda$, $a$, and $M_{\text{DM}}$. Increasing $\Lambda$ raises their real frequencies and damping rates along the displayed trajectories, whereas increasing $M_{\text{DM}}$ has the opposite effect. The dependence on $a$ is nonmonotonic in the damping rate. When $M_{\text{DM}}$ is sufficiently small, the relative shift of the selected $\ell=2$, $n=0$ mode is approximately linear in $M_{\text{DM}}/a$. At fixed $a/M_{\text{B}}=20$, the coefficient fitted by Eq. (\ref{qnm-weak-halo-coefficient-fit}) over the sampled range of $\Lambda$ varies approximately linearly on $\log_{10}(\Lambda M_{\text{B}}^2)$.

There are some things that can be further explored based on such obtained dark matter models. Here, we list some investigations that can be done. Since we only investigate the test scalar QNMs at present, it is more attractive for us to study the QNMs of the gravitational perturbation~\cite{Zhao:2026eti}. Total transmission modes (TTMs) which are another important modes on black hole perturbation theory can also be considered~\cite{Yu:2026rku,Dong:2026zxy,Zhou:2025xdo,Wu:2026hvf,Qian:2025occ,Tuncer:2025dnp}. The rotating counterpart of our model is also very attractive~\cite{Fernandes:2025osu}.

\section*{Acknowledgments}
This work was supported by the National Natural Science Foundation of China under Grants Nos. 12535002, 12588101, 12475063, 12247103, and 12505067, the Ningbo Science \& Technology Bureau under program grant No. 2024DJRC010009, and the National Key R\&D Program of China under Grant No. 2022YFC2204603.

\appendix

\section{Black hole images in various dark matter models}\label{S7}

\subsection{NFW}
\begin{figure}[tbp]
		\begin{minipage}[t]{0.45\linewidth}
			\centering
			\includegraphics[height=4cm]{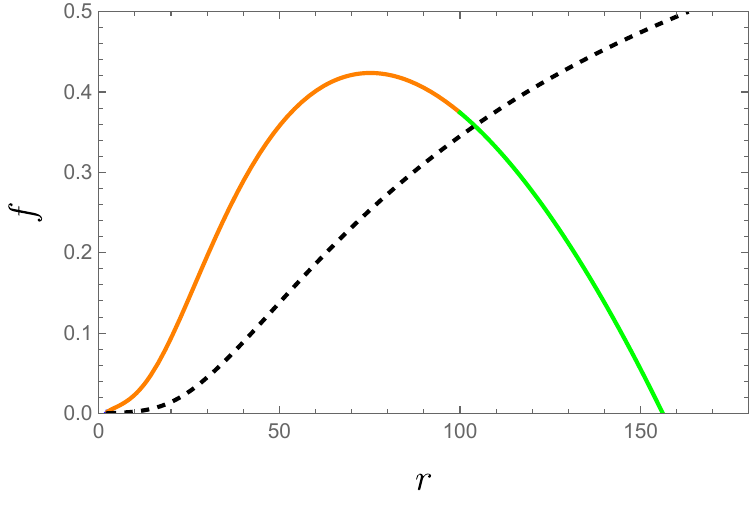}
        \par\vspace{1mm}
		\end{minipage}%
		\begin{minipage}[t]{0.45\linewidth}
			\centering
			\includegraphics[height=4cm]{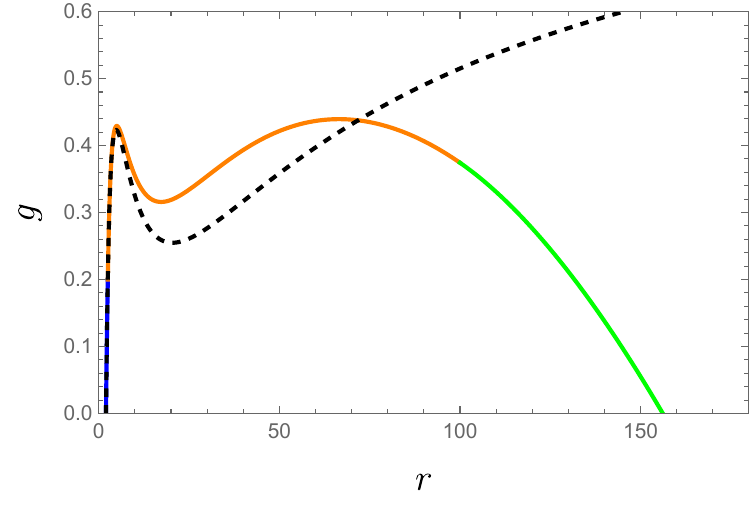}
        \par\vspace{1mm}
		\end{minipage}%

		\begin{minipage}[t]{0.45\linewidth}
			\centering
			\includegraphics[height=4cm]{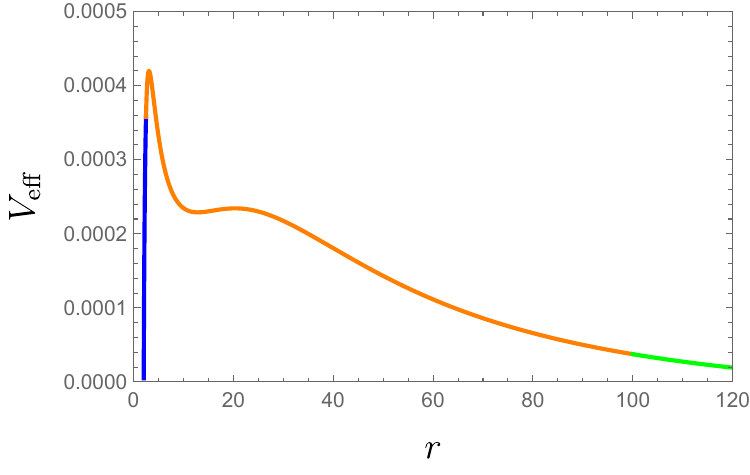}
        \par\vspace{1mm}
		\end{minipage}%
		\begin{minipage}[t]{0.45\linewidth}
			\centering
			\includegraphics[height=4cm]{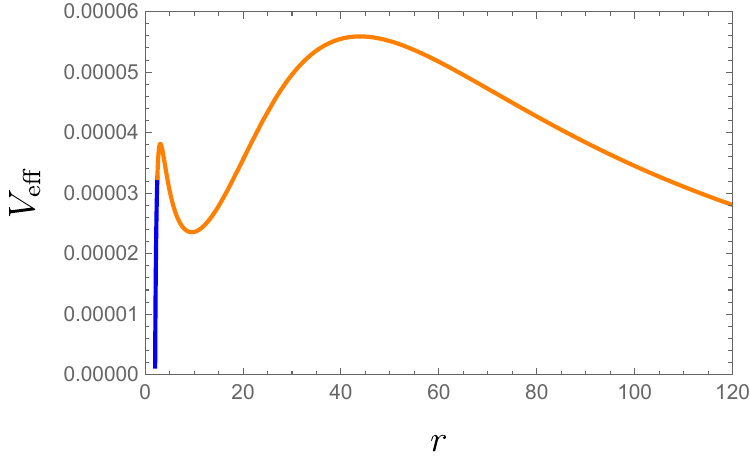}
        \par\vspace{1mm}
		\end{minipage}%
  \caption{NFW-type metric functions $f$ (upper left) and $g$ (upper right), and photon potentials for $\Lambda=10^{-4}M_{\text{B}}^{-2}$ (lower left) and $\Lambda=0$ (lower right). Here $a=6M_{\text{B}}$ and $M_{\text{DM}}=2.8M_{\text{B}}$. The regional colors follow FIG.~\ref{HernV}. Black curves in the upper panels show $\Lambda=0$.}
  \label{NFWV}
\end{figure}

\begin{figure}[tbp]
		\begin{minipage}[t]{0.45\linewidth}
			\centering
			\includegraphics[height=4cm]{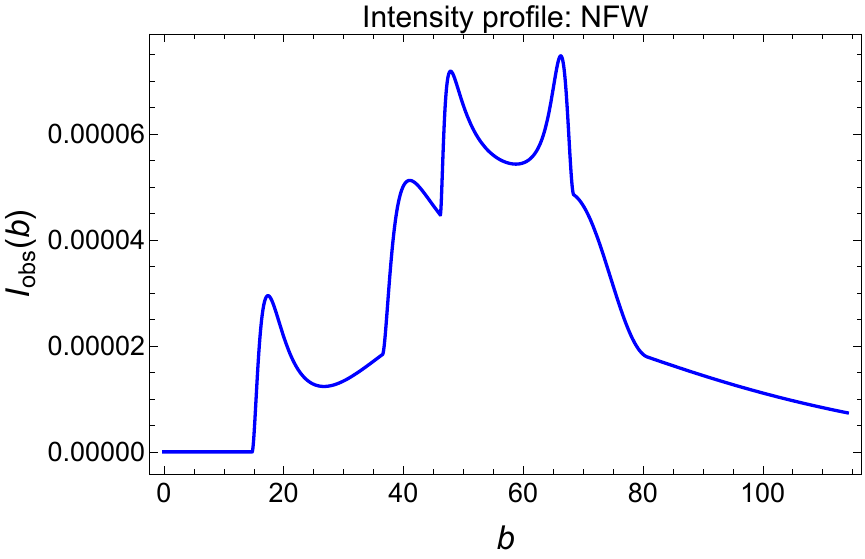}
        \par\vspace{1mm}
		\end{minipage}%
		\begin{minipage}[t]{0.45\linewidth}
			\centering
			\includegraphics[height=4cm]{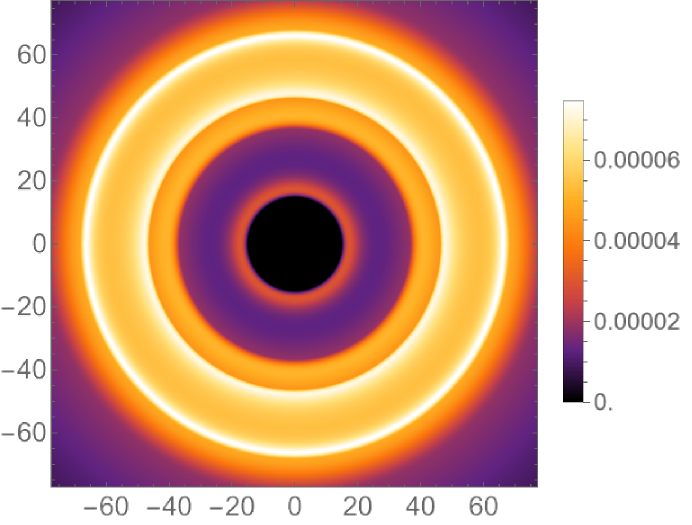}
        \par\vspace{1mm}
		\end{minipage}%

		\begin{minipage}[t]{0.45\linewidth}
			\centering
			\includegraphics[height=4cm]{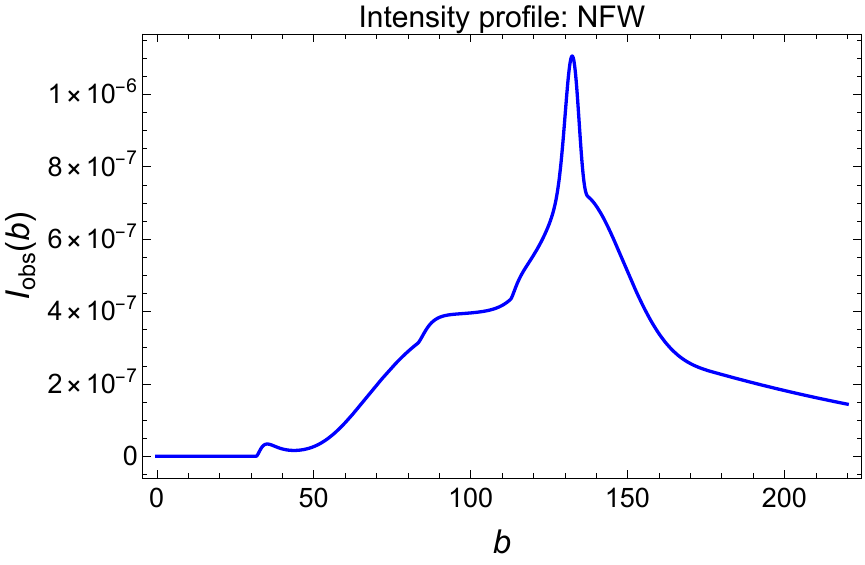}
        \par\vspace{1mm}
		\end{minipage}%
		\begin{minipage}[t]{0.45\linewidth}
			\centering
			\includegraphics[height=4cm]{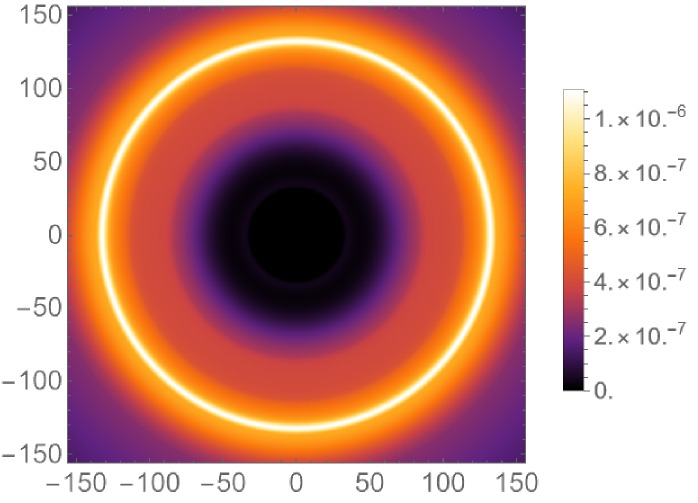}
        \par\vspace{1mm}
		\end{minipage}%
  \caption{NFW-type intensity profiles (left) and images (right) for $a=6M_{\text{B}}$ and $M_{\text{DM}}=2.8M_{\text{B}}$. The upper and lower rows correspond to $\Lambda=10^{-4}M_{\text{B}}^{-2}$ and $\Lambda=0$, respectively.}
  \label{NFW}
\end{figure}

The choice $(\alpha,\gamma,k)=(1,3,1)$ gives the NFW-type model~\cite{Navarro:1994hi}:
\begin{eqnarray}
    \rho(r)&=&\frac{4a\rho_0}{r( 1+r/a)^2} \Big(1-\frac{r_{1}}{r}\Big)\Big(1-\sqrt{\Lambda}r\Big)\, ,\nonumber\\
    m(r)&=&M_{\text{B}}+ \frac{2M_{\text{DM}}}{2\ln2-1} \Big[-\frac{(r-r_{1})}{a+r}+\ln\frac{a+r}{a+r_{1}} \Big]   \nonumber\\
    &&+\sqrt{\Lambda}\frac{2M_{\text{DM}} }{(2\ln2-1)} \Big[-\frac{  (r-r_{1})(r+2a)}{a+r} +(2a+r_{1})\ln\frac{a+r}{a+r_{1}} \Big]\, .
\end{eqnarray}
The mass parameter is defined by the unmodified profile as
\begin{eqnarray}
    M_{\text{DM}}=m(a)|_{M_{\text{B}}=0, \Lambda=0}-M_{\text{B}} =8\pi a^3\rho_0 \left(2\ln2-1 \right) \, .
\end{eqnarray}
Figures~\ref{NFWV} and \ref{NFW} show the metric, photon potential, and images for $a=6M_{\text{B}}$ and $M_{\text{DM}}=2.8M_{\text{B}}$. As in the Hernquist example, the positive-$\Lambda$ image has a bright band associated with two photon spheres. It is brighter and has a smaller central dark region and ring scale than the $\Lambda=0$ image. In the latter case, the higher outer barrier leaves only one prominent ring.

\subsection{Jaffe}
\begin{figure}[tbp]
		\begin{minipage}[t]{0.45\linewidth}
			\centering
			\includegraphics[height=4cm]{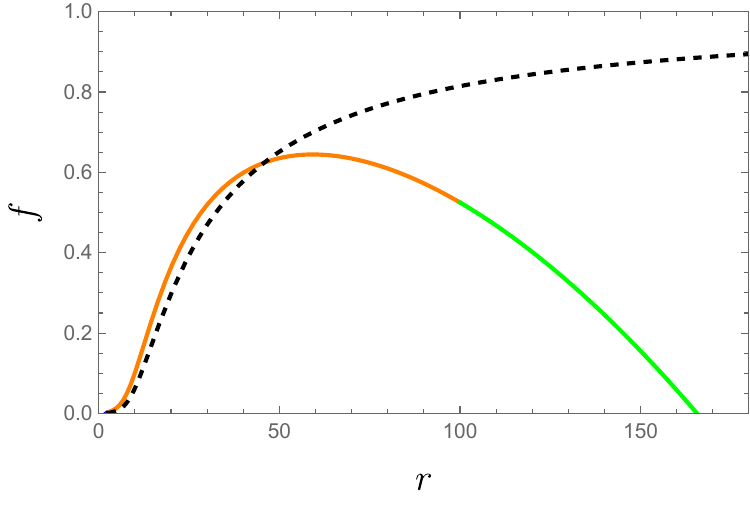}
        \par\vspace{1mm}
		\end{minipage}%
		\begin{minipage}[t]{0.45\linewidth}
			\centering
			\includegraphics[height=4cm]{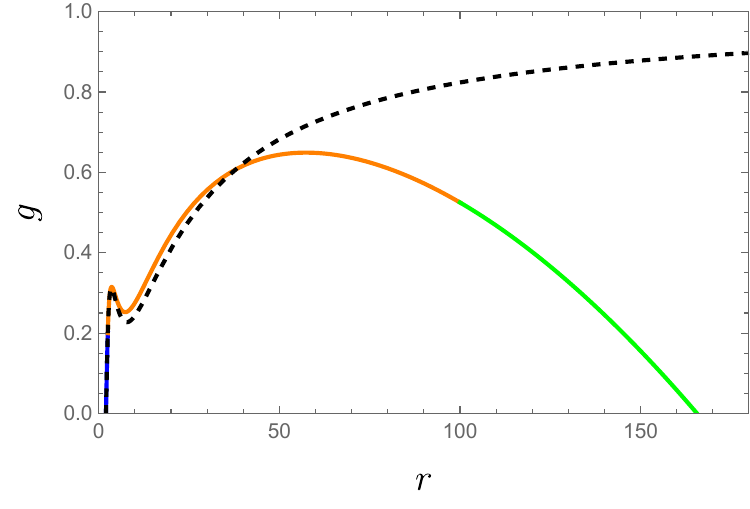}
        \par\vspace{1mm}
		\end{minipage}%

		\begin{minipage}[t]{0.45\linewidth}
			\centering
			\includegraphics[height=4cm]{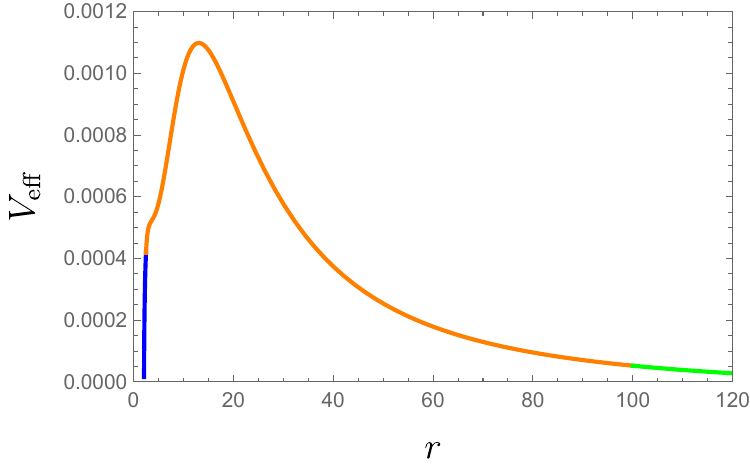}
        \par\vspace{1mm}
		\end{minipage}%
		\begin{minipage}[t]{0.45\linewidth}
			\centering
			\includegraphics[height=4cm]{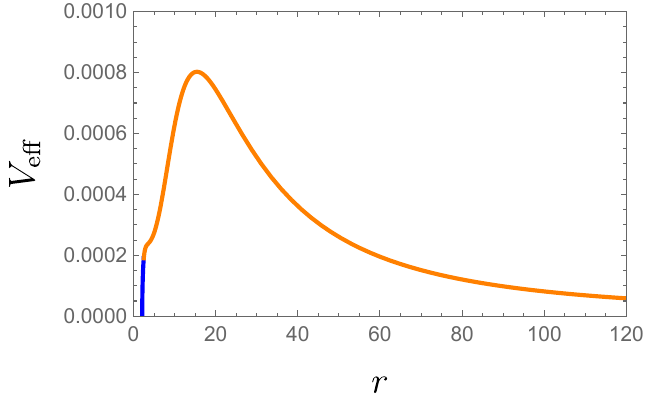}
        \par\vspace{1mm}
		\end{minipage}%
  \caption{Jaffe-type metric functions $f$ (upper left) and $g$ (upper right), and photon potentials for $\Lambda=10^{-4}M_{\text{B}}^{-2}$ (lower left) and $\Lambda=0$ (lower right). Here $a=6M_{\text{B}}$ and $M_{\text{DM}}=18M_{\text{B}}$. Colors follow Figs.~\ref{Hernfg} and \ref{HernV}.}
  \label{JaffeV}
\end{figure}

\begin{figure}[tbp]
		\begin{minipage}[t]{0.45\linewidth}
			\centering
			\includegraphics[height=4cm]{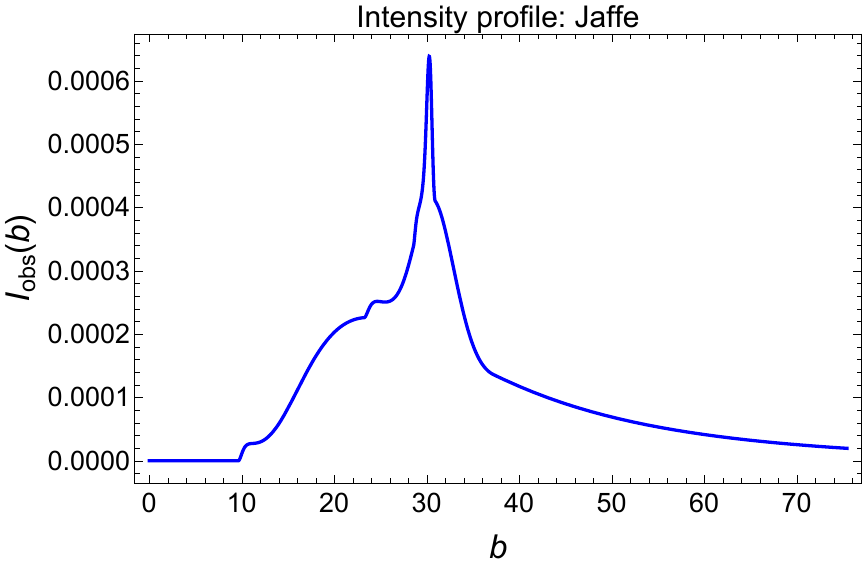}
        \par\vspace{1mm}
		\end{minipage}%
		\begin{minipage}[t]{0.45\linewidth}
			\centering
			\includegraphics[height=4cm]{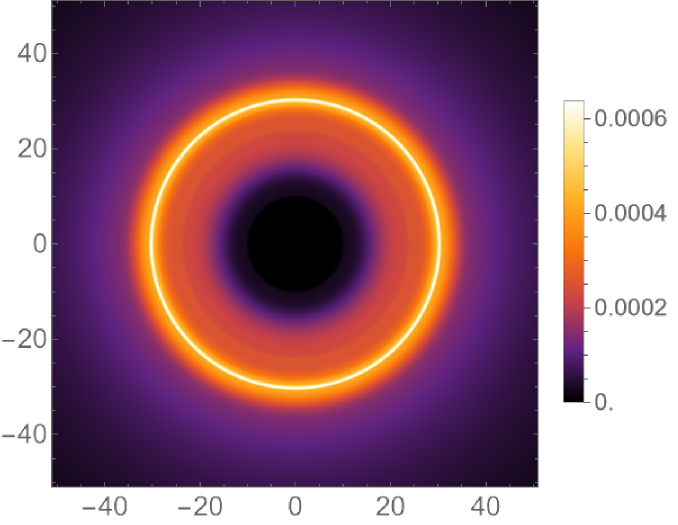}
        \par\vspace{1mm}
		\end{minipage}%

		\begin{minipage}[t]{0.45\linewidth}
			\centering
			\includegraphics[height=4cm]{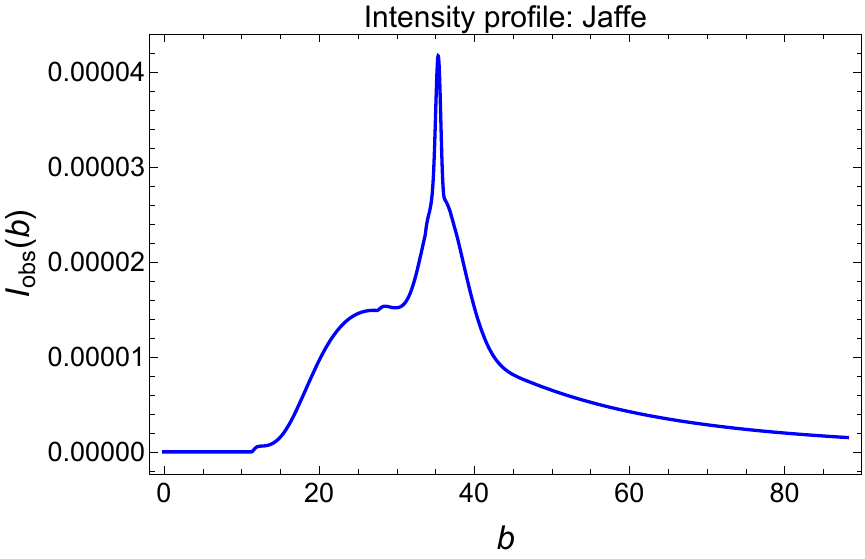}
        \par\vspace{1mm}
		\end{minipage}%
		\begin{minipage}[t]{0.45\linewidth}
			\centering
			\includegraphics[height=4cm]{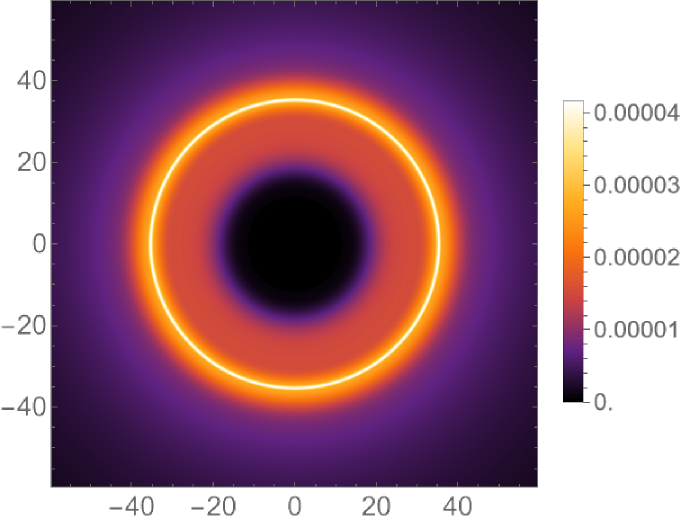}
        \par\vspace{1mm}
		\end{minipage}%
  \caption{Jaffe-type intensity profiles (left) and images (right) for $a=6M_{\text{B}}$ and $M_{\text{DM}}=18M_{\text{B}}$. The upper and lower rows correspond to $\Lambda=10^{-4}M_{\text{B}}^{-2}$ and $\Lambda=0$, respectively.}
  \label{Jaffe}
\end{figure}

For $(\alpha,\gamma,k)=(2,4,1)$, the Jaffe-type model~\cite{Jaffe:1983iv} has
\begin{eqnarray}
    \rho(r)&=&\frac{4a^2\rho_0}{r^2( 1+r/a)^2} \left(1-\frac{r_{\text{1}}}{r}\right)\Big(1-\frac{r}{r_0}\Big)     \, ,\nonumber\\
    m(r)&=&M_{\text{B}}+ \frac{M_{\text{DM}}}{a} \Big[\frac{a(r-r_{1})}{a+r} -r_{1}\ln\frac{r}{r_{1}}+r_{1}\ln\frac{a+r}{a+r_{1}} \Big] \,\nonumber\\
    &&+\sqrt{\Lambda}\frac{M_{\text{DM}} }{a} \Big[\frac{a^2(r-r_{1}) }{a+r} -a^2\ln\frac{a+r}{a+r_{1}} \Big]     \, .
\end{eqnarray}
with
\begin{eqnarray}
    M_{\text{DM}}=m(+\infty)|_{M_{\text{B}}=0, \Lambda=0}-M_{\text{B}} =16\pi a^3 \rho_0 \, .
\end{eqnarray}
Within the allowed parameter range, some configurations have two photon spheres. In the cases examined, however, their critical impact parameters are too close to produce clearly separated image features. Figures~\ref{JaffeV} and \ref{Jaffe} instead show a single-photon-sphere example with $a=6M_{\text{B}}$ and $M_{\text{DM}}=18M_{\text{B}}$. Its images have similar morphology with and without $\Lambda$. The main difference is their brightness.

\subsection{Burkert}
\begin{figure}[tbp]
		\begin{minipage}[t]{0.45\linewidth}
			\centering
			\includegraphics[height=4cm]{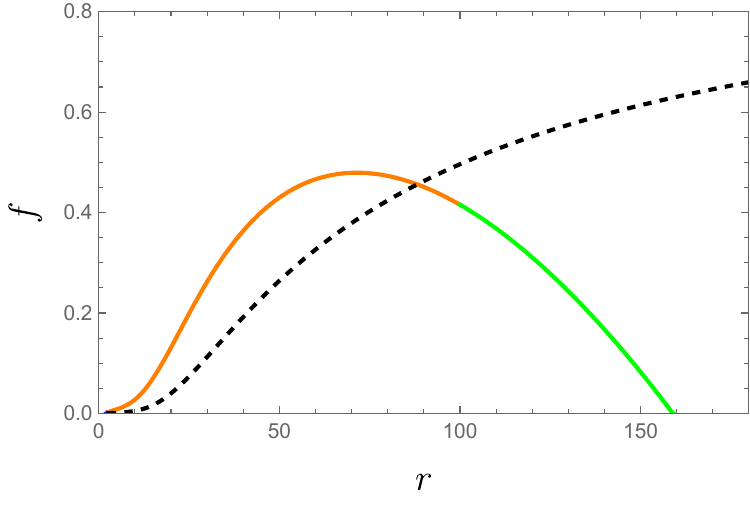}
        \par\vspace{1mm}
		\end{minipage}%
		\begin{minipage}[t]{0.45\linewidth}
			\centering
			\includegraphics[height=4cm]{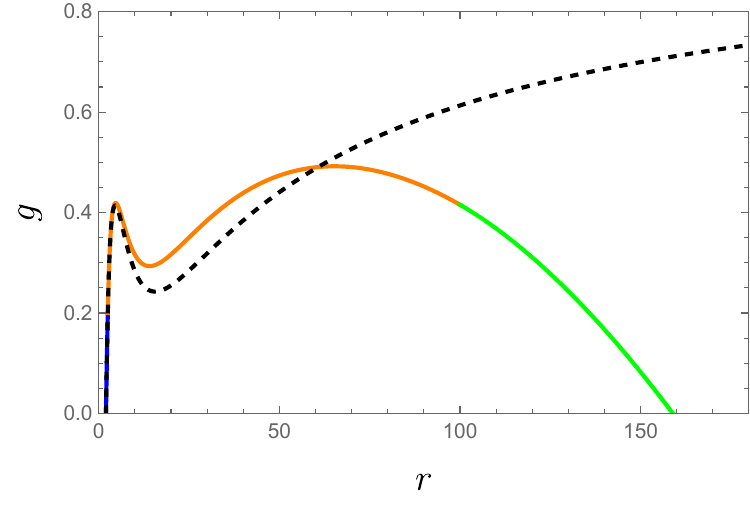}
        \par\vspace{1mm}
		\end{minipage}%

		\begin{minipage}[t]{0.45\linewidth}
			\centering
			\includegraphics[height=4cm]{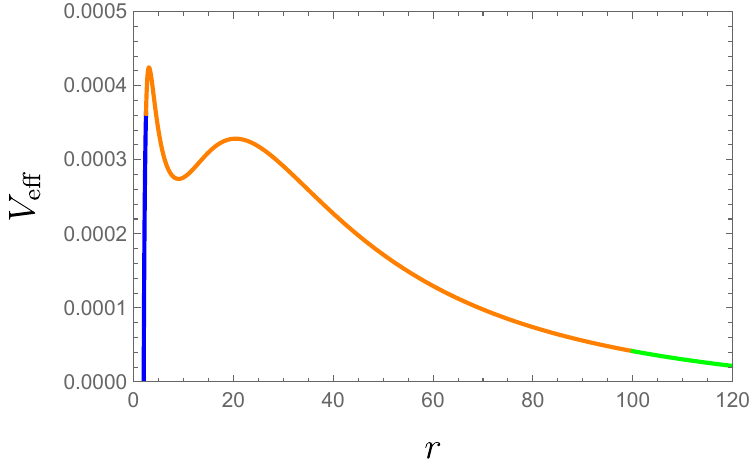}
        \par\vspace{1mm}
		\end{minipage}%
		\begin{minipage}[t]{0.45\linewidth}
			\centering
			\includegraphics[height=4cm]{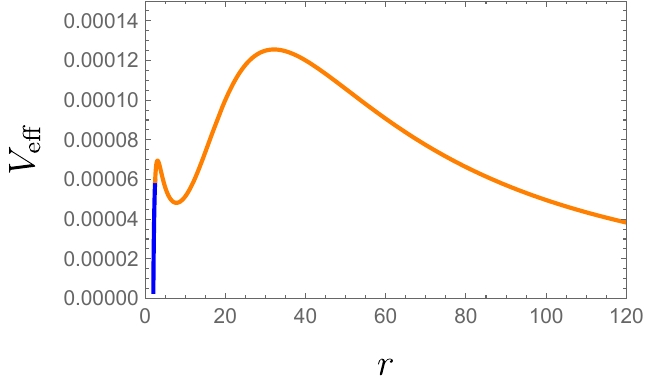}
        \par\vspace{1mm}
		\end{minipage}%
  \caption{Burkert-model metric functions $f$ (upper left) and $g$ (upper right), and photon potentials for $\Lambda=10^{-4}M_{\text{B}}^{-2}$ (lower left) and $\Lambda=0$ (lower right). Here $a=6M_{\text{B}}$ and $M_{\text{DM}}=2.8M_{\text{B}}$. Colors follow Figs.~\ref{Hernfg} and \ref{HernV}.}
  \label{BurV}
\end{figure}

\begin{figure}[tbp]
		\begin{minipage}[t]{0.45\linewidth}
			\centering
			\includegraphics[height=4cm]{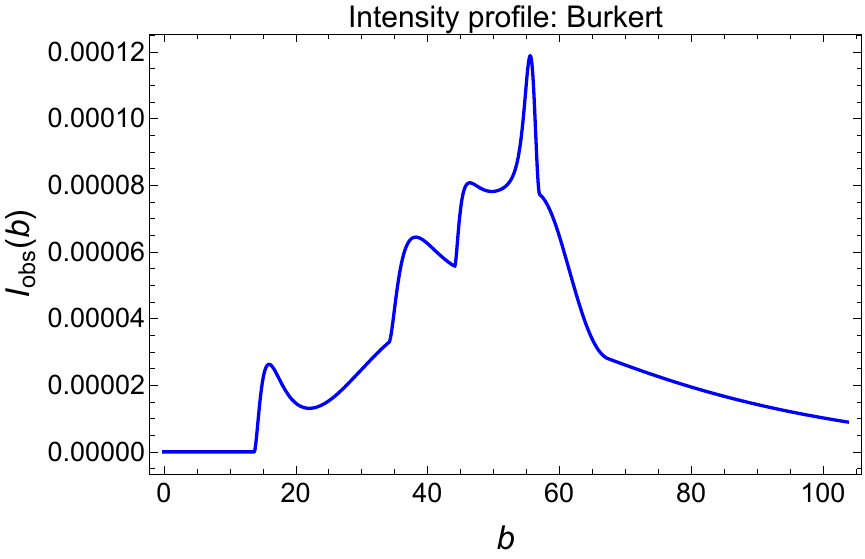}
        \par\vspace{1mm}
		\end{minipage}%
		\begin{minipage}[t]{0.45\linewidth}
			\centering
			\includegraphics[height=4cm]{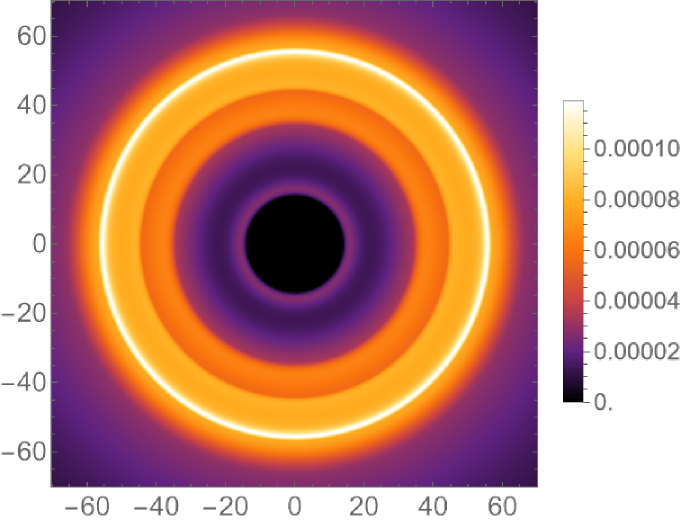}
        \par\vspace{1mm}
		\end{minipage}%

		\begin{minipage}[t]{0.45\linewidth}
			\centering
			\includegraphics[height=4cm]{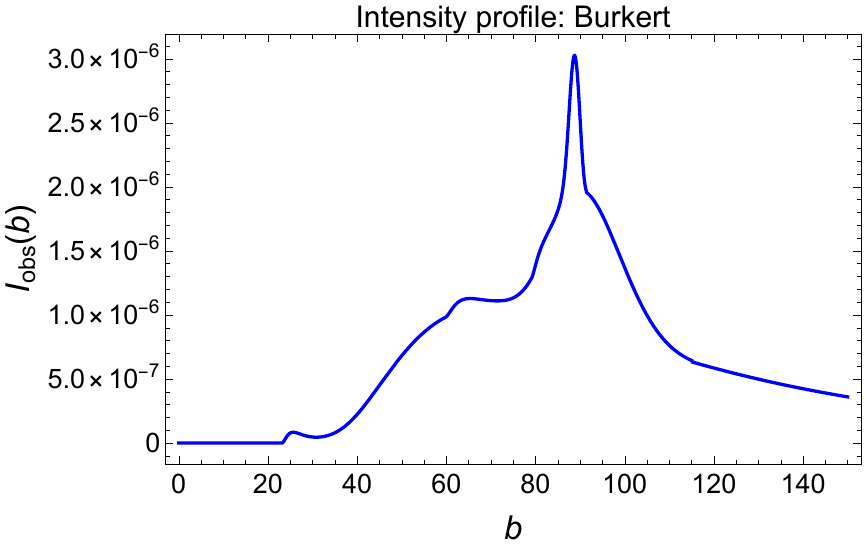}
        \par\vspace{1mm}
		\end{minipage}%
		\begin{minipage}[t]{0.45\linewidth}
			\centering
			\includegraphics[height=4cm]{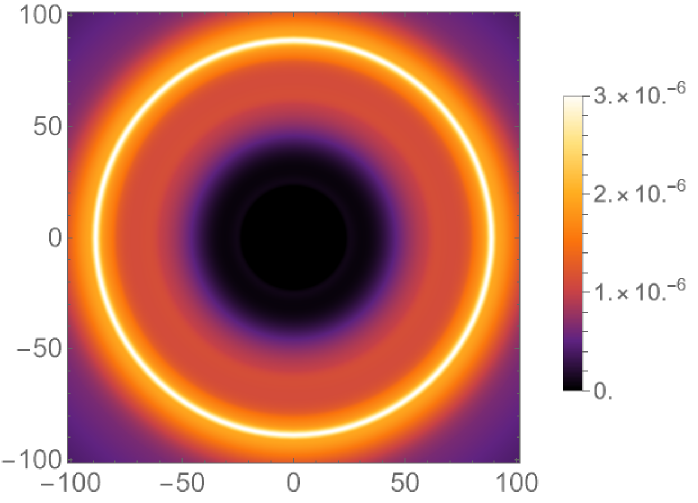}
        \par\vspace{1mm}
		\end{minipage}%
  \caption{Burkert-model intensity profiles (left) and images (right) for $a=6M_{\text{B}}$ and $M_{\text{DM}}=2.8M_{\text{B}}$. The upper and lower rows correspond to $\Lambda=10^{-4}M_{\text{B}}^{-2}$ and $\Lambda=0$, respectively.}
  \label{Burkert}
\end{figure}

For the model denoted here as Burkert, we use $(\alpha,\gamma,k)=(1,3,2)$~\cite{Burkert:1995yz,Salucci:2000ps}:
\begin{eqnarray}
    \rho(r)&=&\frac{2a\rho_0}{r( 1+r^2/a^2)}     \Big(1-\frac{r_{1}}{r}\Big)\Big(1-\frac{r}{r_0}\Big)     \, ,\nonumber\\
    m(r)&=&M_{\text{B}}-\frac{2M_{\text{DM}} }{\ln2} \Big[\frac{r_{1}}{a}\Big(\arctan\frac{r}{a}-\arctan\frac{r_{1}}{a}\Big)-\frac{1}{2}\ln\frac{a^2+r^2}{a^2+r_{1}^2}\Big]  \nonumber\\
    &&-\sqrt{\Lambda}\frac{2M_{\text{DM}} }{\ln2} \Big[(r-r_{1})  -a\Big(\arctan\frac{r}{a}-\arctan\frac{r_{1}}{a}\Big)-\frac{1}{2}r_{1}\ln\frac{a^2+r^2}{a^2+r_{1}^2}\Big]\, .
\end{eqnarray}
where
\begin{eqnarray}
    M_{\text{DM}}=m(a)|_{M_{\text{B}}=0, \Lambda=0}-M_{\text{B}} =4\ln2\pi a^3 \rho_0 \, .
\end{eqnarray}
Figures~\ref{BurV} and \ref{Burkert} show the results for $a=6M_{\text{B}}$ and $M_{\text{DM}}=2.8M_{\text{B}}$. The image comparison resembles the Hernquist case: in the chosen setup, $\Lambda>0$ produces a brighter image with a smaller ring or band and a smaller central dark region.

\subsection{Taylor--Silk}
\begin{figure}[tbp]
		\begin{minipage}[t]{0.45\linewidth}
			\centering
			\includegraphics[height=4cm]{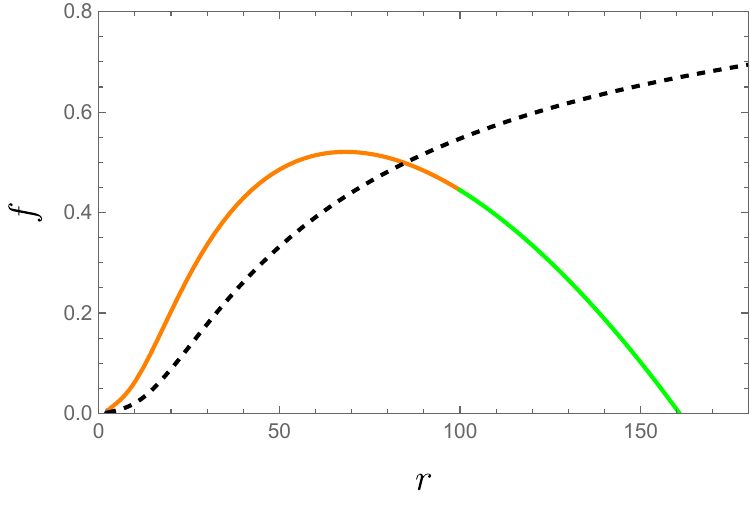}
        \par\vspace{1mm}
		\end{minipage}%
		\begin{minipage}[t]{0.45\linewidth}
			\centering
			\includegraphics[height=4cm]{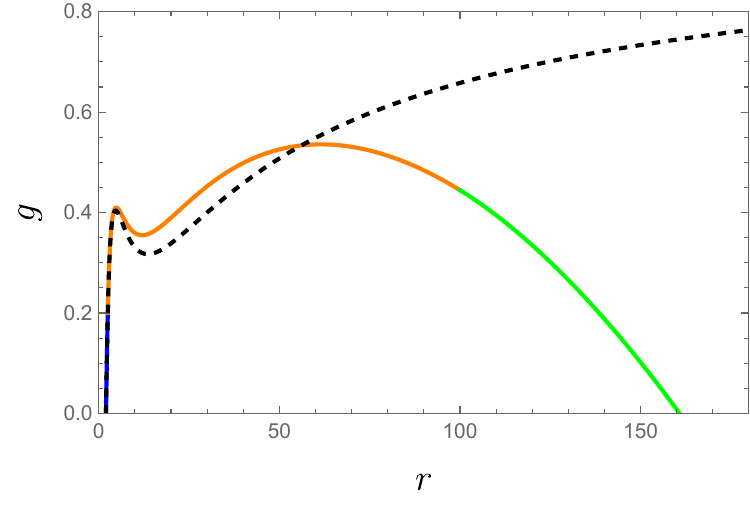}
        \par\vspace{1mm}
		\end{minipage}%

		\begin{minipage}[t]{0.45\linewidth}
			\centering
			\includegraphics[height=4cm]{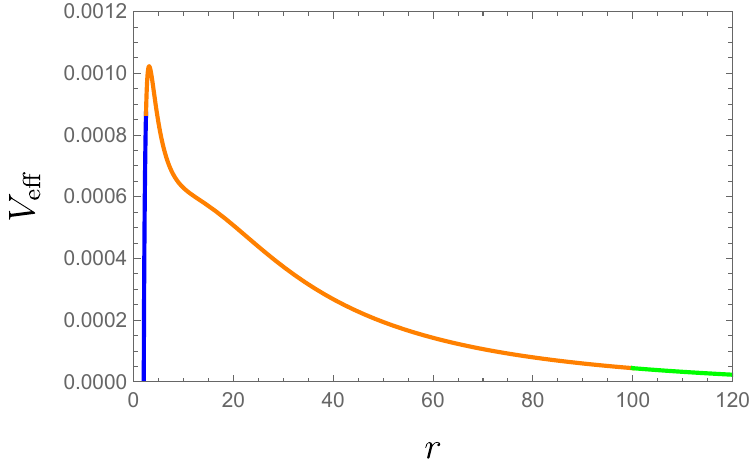}
        \par\vspace{1mm}
		\end{minipage}%
		\begin{minipage}[t]{0.45\linewidth}
			\centering
			\includegraphics[height=4cm]{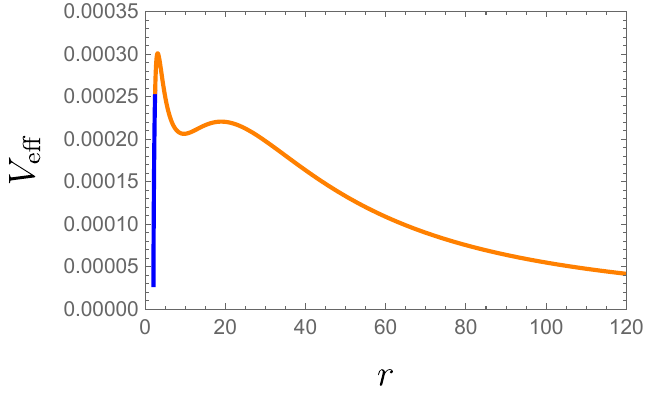}
        \par\vspace{1mm}
		\end{minipage}%
  \caption{Taylor--Silk-type metric functions $f$ (upper left) and $g$ (upper right), and photon potentials for $\Lambda=10^{-4}M_{\text{B}}^{-2}$ (lower left) and $\Lambda=0$ (lower right). Here $a=6M_{\text{B}}$ and $M_{\text{DM}}=3.4M_{\text{B}}$. Colors follow Figs.~\ref{Hernfg} and \ref{HernV}.}
  \label{TSV}
\end{figure}

\begin{figure}[tbp]
		\begin{minipage}[t]{0.45\linewidth}
			\centering
			\includegraphics[height=4cm]{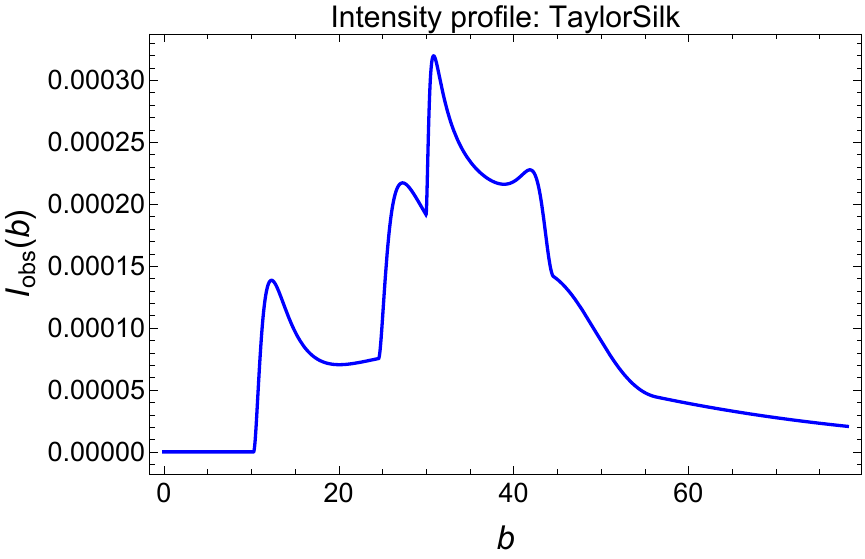}
        \par\vspace{1mm}
		\end{minipage}%
		\begin{minipage}[t]{0.45\linewidth}
			\centering
			\includegraphics[height=4cm]{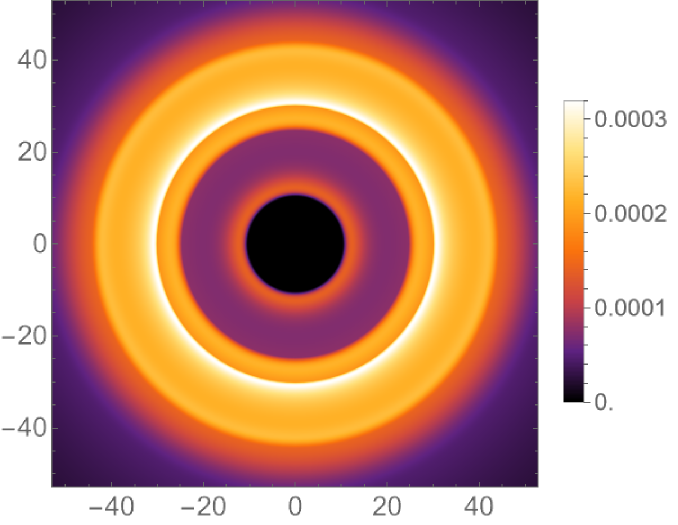}
        \par\vspace{1mm}
		\end{minipage}%

		\begin{minipage}[t]{0.45\linewidth}
			\centering
			\includegraphics[height=4cm]{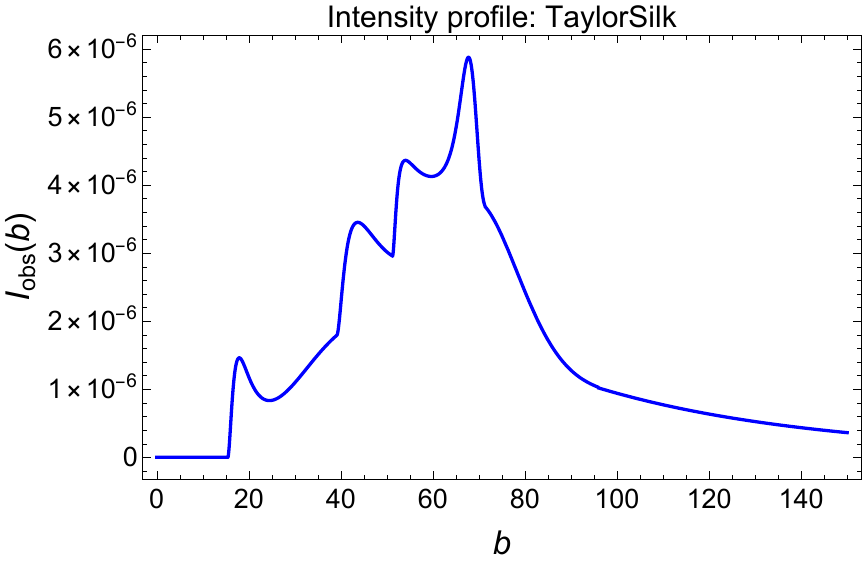}
        \par\vspace{1mm}
		\end{minipage}%
		\begin{minipage}[t]{0.45\linewidth}
			\centering
			\includegraphics[height=4cm]{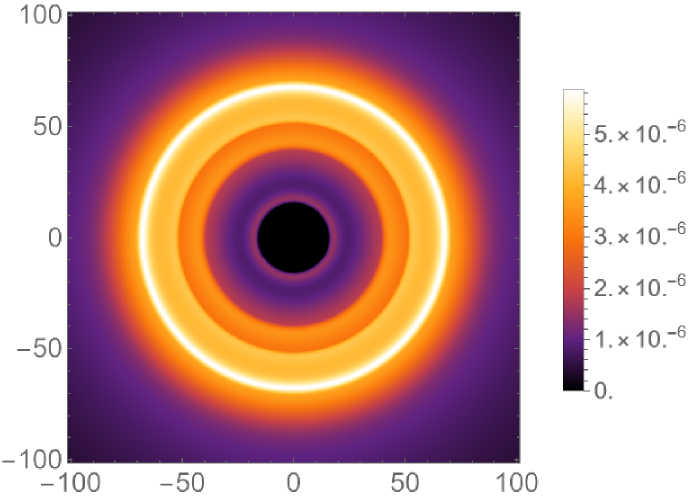}
        \par\vspace{1mm}
		\end{minipage}%
  \caption{Taylor--Silk-type intensity profiles (left) and images (right) for $a=6M_{\text{B}}$ and $M_{\text{DM}}=3.4M_{\text{B}}$. The upper and lower rows correspond to $\Lambda=10^{-4}M_{\text{B}}^{-2}$ and $\Lambda=0$, respectively.}
  \label{Taylorsilk}
\end{figure}

The Taylor--Silk-type model~\cite{Taylor:2002zd}, with $(\alpha,\gamma,k)=(3/2,3,3/2)$, is
\begin{eqnarray}
    \rho(r)&=&\frac{2a^\frac{3}{2}\rho_0}{r^\frac{3}{2}\big[ 1+(r/a)^\frac{3}{2} \big]}      \Big(1-\frac{r_{\text{1}}}{r}\Big)\Big(1-\frac{r}{r_0}\Big) \, ,\nonumber\\
    m(r)&=&M_{\text{B}} -\frac{M_{\text{DM}} }{2\ln2 a} \Bigg[ -2\sqrt{3}r_{\text{1}}\Big(\arctan\frac{1-2\sqrt{r/a}}{\sqrt{3}}-\arctan\frac{1-2\sqrt{r_1/a}}{\sqrt{3}}\Big)\nonumber\\
    &&+r_{1} \Big(2\ln\frac{\sqrt{a}+\sqrt{r}}{\sqrt{a}+\sqrt{r_{1}}}-\ln\frac{a+r-\sqrt{ar}}{a+r_{1}-\sqrt{ar_{1}}}  \Big)
    -2a\ln\frac{a^\frac{3}{2}+r^\frac{3}{2}}{a^\frac{3}{2}+r_{1}^\frac{3}{2}}  \Bigg] \nonumber\\
    &&-\sqrt{\Lambda}\frac{M_{\text{DM}} }{2\ln2 a} \Bigg[ 3a(r-r_{1}) + 2\sqrt{3}a^2\Big(\arctan\frac{1-2\sqrt{r/a}}{\sqrt{3}}-\arctan\frac{1-2\sqrt{r_1/a}}{\sqrt{3}}\Big) \nonumber\\
    &&+a^2 \Big(2\ln\frac{\sqrt{a}+\sqrt{r}}{\sqrt{a}+\sqrt{r_{1}}}-\ln\frac{a+r-\sqrt{ar}}{a+r_{1}-\sqrt{ar_{1}}}  \Big)
    -2ar_{1}\ln\frac{a^\frac{3}{2}+r^\frac{3}{2}}{a^\frac{3}{2}+r_{1}^\frac{3}{2}}  \Bigg] \, .
\end{eqnarray}
with
\begin{eqnarray}
    M_{\text{DM}}=m(a)|_{M_{\text{B}}=0, \Lambda=0}-M_{\text{B}} =\frac{16\ln2}{3}\pi a^3\rho_0 \, .
\end{eqnarray}
Figures~\ref{TSV} and \ref{Taylorsilk} use $a=6M_{\text{B}}$ and $M_{\text{DM}}=3.4M_{\text{B}}$. The $\Lambda=0$ potential has two photon-sphere maxima, whereas the positive-$\Lambda$ potential has only one. Nevertheless, both images contain a bright band and differ mainly in its scale and brightness.

In the positive-$\Lambda$ example, the band derive from the emission and the redshift factor rather than with two photon spheres. The peak of the redshift factor is displaced from the region of strongest emission, so the observed intensity need not peak at the same radius as the emitted intensity. A broad bright band is therefore not, by itself, evidence for two photon spheres.

\end{document}